\documentclass[10pt]{article}
\usepackage{wrapfig,array,ulem}
\usepackage{epsfig,cancel,amsthm,amssymb}
\usepackage{color,tikz}
\usetikzlibrary{decorations.markings,arrows, calc}
\usepackage{color,todonotes}
\usepackage{graphicx,framed,verbatim,caption}
\usepackage[colorlinks=true, pdfstartview=FitV, linkcolor=darkblue, citecolor=darkblue, urlcolor=darkblue]{hyperref}
\definecolor{shadecolor}{rgb}{0.9, 0.9, 0.86}
\definecolor{darkgreen}{rgb}{0.2, 0.5,  0}
\definecolor{darkblue}{rgb}{0.1,0.1,0.45}
\def\RR{ {\mathbf R} }

\def\&{\vspace{-5pt}&}

\def\Re{\mathrm {Re}\,}

\def\J{{\mathbf J}}

\def\A{{\mathbf A}}

\def \AA {\mathbf {Ai}}

\def \eqref#1{(\ref{#1})}
\def \& {&\hspace{-10pt}}
\def\Ai{ {\mathrm {Ai}}}

\def \wt{\widetilde}
\def\t{ {\mathbf t}}
\newcommand{\G}{\Gamma} 
\renewcommand{\d}{\mathrm d}
\newcommand{\pa}{\partial}       
\newtheorem{theorem}{Theorem}[section]
\newtheorem{example}[theorem]{Example}
\newtheorem{exercise}[theorem]{Exercise}

\newtheorem{lemma}[theorem]{Lemma}
\newtheorem{remark}[theorem]{Remark}
\newtheorem{problem}[theorem]{Riemann-Hilbert Problem}

\newtheorem{proposition}[theorem]{Proposition} 
\newtheorem{corollary}[theorem]{Corollary} 
\newtheorem{question}[theorem]{Question} 
\newtheorem{definition}[theorem]{Definition}

\def\le{\left}
\def\ri{\right}
\def\ds{\displaystyle}

\def\res{\mathop{\mathrm {res}}\limits_}

\def\bt{\begin{theorem}}
\def\et{\end{theorem}}
\def\bc{\begin{corollary}}
\def\ec{\end{corollary}}
\def\bx{\begin{example}}
\def\ex{\end{example}}
\def\bxr{\begin{exercise}\small}
\def\exr{\end{exercise}}
\def\bl{\begin{lemma}}
\def\el{\end{lemma}}
\def\bd{\begin{definition}}
\def\ed{\end{definition}}
\def\bp{\begin{proposition}}
\def\ep{\end{proposition}}

\def\br{\begin{remark}}
\def\er{\end{remark}}

\def\be{\begin{eqnarray}}
\def\ee{\end{eqnarray}}
\def \Tr {\mathrm{Tr}\,}
\def\&{\hspace{-15pt}&}
\def\bea{\begin{eqnarray}}
\def\eea{\end{eqnarray}}
\def\beas{\begin{eqnarray*}}
\def\eeas{\end{eqnarray*}}

\def \pa{\partial}
\def\C{{\mathbb C}}
\def\L{\mathcal L}
\def\R{{\mathbb R}}
\def\N{{\mathbb N}}

\def\H{{\cal H}}
\def\Z{{\mathbb Z}}

\def\l{ \lambda }
\def\1{{\bf 1}}
\def\s{ {\sigma}} 
\def\z{\zeta}

\def\QED {\hfill $\blacksquare$\par\vskip 3pt}

\renewcommand{\theequation}{\arabic{section}.\arabic{equation}}

\makeatletter
\@addtoreset{equation}{section}
\makeatother

\begin{document}

\baselineskip 14pt plus 1pt minus 1pt

\begin{flushright}
\end{flushright}
\vspace{0.2cm}
\begin{center}
\begin{Large}
\textbf{ 
 The   Kontsevich matrix integral: convergence to the  Painlev\'e\ hierarchy and Stokes' phenomenon
} 
\end{Large}
\end{center}
\bigskip
\begin{center}
M. Bertola$^{\dagger\ddagger \clubsuit}$\footnote{Marco.Bertola@\{concordia.ca, sissa.it\}},  
M. Cafasso $^{\diamondsuit}$ \footnote{cafasso@math.univ-angers.fr}.
\\
\bigskip
\begin{minipage}{0.7\textwidth}
\begin{small}
\begin{enumerate}
\item [${\dagger}$] {\it  Department of Mathematics and
Statistics, Concordia University\\ 1455 de Maisonneuve W., Montr\'eal, Qu\'ebec,
Canada H3G 1M8} 
\item[${\ddagger}$] {\it SISSA/ISAS, via Bonomea 265, Trieste, Italy }
\item[${\clubsuit}$] {\it Centre de recherches math\'ematiques,
Universit\'e de Montr\'eal\\ C.~P.~6128, succ. centre ville, Montr\'eal,
Qu\'ebec, Canada H3C 3J7} 
\item [${\diamondsuit}$] {\it LAREMA, Universit\'e d'Angers\\ 2 Boulevard Lavoisier, 49045 Angers, France.}
\end{enumerate}
\end{small}
\end{minipage}
\vspace{0.5cm}
\end{center}
\bigskip
\begin{center}
\begin{abstract}
We show that the Kontsevich integral on $n\times n$ matrices ($n< \infty$) is the isomonodromic tau function associated to a $2\times 2$ Riemann--Hilbert problem.
The approach allows us to gain control of the analysis of the convergence as $n\to\infty$. By an appropriate choice of the external source matrix  in  Kontsevich's integral, we show that the limit produces the isomonodromic tau function of a special tronqu\'ee solution of the first Painlev\'e\ hierarchy, and we identify the solution in terms of the Stokes' data of the associated linear problem. We also show that there are several tau functions that are analytic in appropriate sectors of the space of parameters and that the formal Witten-Kontsevich tau function is  the asymptotic expansion of each of them in their respective sectors, thus providing an analytic tool  to analyze its  nonlinear Stokes' phenomenon. 
\end{abstract}
\end{center}
\tableofcontents

\section{Introduction and results}
The Kontsevich matrix integral has been introduced in \cite{Kontsevich:1992p30} as a tool to prove the Witten conjecture, which relates the intersection numbers of the Deligne--Mumford moduli space to a specific solution of the Korteweg--de Vries hierarchy. This integral  is given by the following expression\footnote{ We normalize the variables of integration differently from \cite{Kontsevich:1992p30}. See Rem. \ref{normalisation} for the precise comparison.}
\bea
\label{ZnK}
Z_n(x; {Y}):=\frac{ \ds \int_{H_n} \d M {\rm e}^{ \Tr \le(i\frac {M^3}3 - {Y} M^2 + i x M\ri)}}{ \int_{H_n} \d M {\rm e}^{- \Tr \le( {Y} M^2\ri)}},
\eea
where the integral is over the space $H_n$ of $(n\times n)$ Hermitian matrices and $Y$ is a diagonal matrix whose entries $y_k, \ k= 1,\ldots,n$ satisfy the condition $\Re y_k>0$ (to ensure convergence); the parameter $x$ was absent in the original formulation and  it is added here for later convenience.
 Kontsevich proved that the function $Z_n( 0; Y )$ is  a ratio of the Wronskian of Airy functions and the Vandermonde determinant of the eigenvalues of $ {Y}$:
 \be
 \label{ZnKY}
 Z_n(x;{Y}) = 2^n \pi^\frac n2 {\rm e}^{\frac 2 3 \Tr  {Y}^3   + x \Tr Y}\frac{\det\le[\Ai^{(j-1)} (y_k^2 + x ) \ri]_{k,j\leq n} \prod_{j=1}^n(y_j)^{\frac 1 2} }{\prod_{j<k} (y_j-y_k) }, \ \ \ \Re y_j>0.
 \ee
(See App. \ref{proofZnKY} for a simple proof).
A closely related model is the   {\it external source} matrix model,
with a probability measure of the form 
\be
\d\mu(M) \propto  {\rm e}^{\Tr(V(M) + \Lambda M)} \d M,
\ee
where $\Lambda = {\rm diag}(\l_1,\dots, \l_n)$, 
 where, in this context, the function $V(x)$ is a real--valued scalar function. If one considers it as a  random matrix model for the eigenvalues of $M$ then the usual approach of orthogonal polynomials \cite{Deift} needs to be generalized to multi-orthogonal polynomials. Then the familiar $2\times 2$ Riemann--Hilbert problem for the orthogonal polynomials trades places with a different Riemann Hilbert problem of size  $r\times r$, where $r$ is the number of distinct eigenvalues of the matrix $ {Y}$ and the orthogonality is replaced by multiple orthogonality \cite{Aptekarev_etal_ExtSource}.
In general (except for special cases \cite{ClaeysWang14}), the case with $n$ distinct eigenvalues leads naturally  to a Riemann--Hilbert problem of size $n+1$. Our goals are  however  different: we are interested in the integral \eqref{ZnK} itself and to study rigorously its limit as $n\to \infty$ and its convergence to particular tau functions of the first Painlev\'e\ hierarchy.\\

The equation \eqref{ZnKY} is the key step to prove that the Kontsevich integral is a tau function (in the formal sense of Sato \cite{SatoKP}) for the KdV hierarchy, where the eigenvalues $y_k$ plays the role of Miwa variables (see eq. \eqref{Miwa}).  The  first goal  of this paper is to identify the Kontsevich integral with another type of tau function, of the type introduced by Jimbo, Miwa and Ueno \cite{JMU1,JMU2} in the study of isomonodromic deformations of linear ODEs; the so called isomonodromic tau function.

Our approach is conceptually equivalent to the following: consider the ``bare system'' 
\be
\frac {\d}{\d \l} \Psi_0 (\l;x) = \le[ \begin{array}{cc}
0 & -i\\
i(\l + x) & 0  
\end{array}\ri]\Psi_0(\l;x) 
\qquad \frac {\d}{\d x} \Psi_0 (\l;x) = \le[ \begin{array}{cc}
0 &-i\\
i(\l + x) & 0  
\end{array}\ri]\Psi_0(\l;x).
\label{14}
\ee

A fundamental matrix joint solution of \eqref{14} (up to right multiplication by an invertible matrix) can be written explicitly in terms of Airy functions (Section \ref{sectionbare}).
We then proceed with a ``dressing'',
 namely, a sequence of $n$ discrete Schlesinger transformations (in the sense of \cite{JMU2}) in which the monodromy data (Stokes' matrices) are preserved but we allow $\Psi_n$ to have $n$ poles at the points $\{\lambda_1,\ldots,\lambda_n\} := \vec \l$ with  $\l_k = y_k^2,\ k =1, \ldots,n$. 
%
The result of this operation is a  system of partial differential equations for the unknown matrix valued function $\Psi_n$ of the form
\be
\frac {\pa}{\pa \l}\Psi_n(\l; x, \vec \l )  &\& = A(\l;x, \vec \l   ) \Psi _n(\l; x,\vec \l ) \label{psil}\\
\frac {\pa}{\pa x}\Psi_n  (\l;x, \vec \l   ) &\& = U(\l;x,\vec \l  ) \Psi_n (\l;x, \vec \l  ) \label{psix}\\
\frac{\pa}{\pa \l_k}\Psi_n (\l;x, \vec \l   ) &\&= -\frac{A_k (x ,\vec \l  )}{\l-\l_k}  \Psi_n(\l;x, \vec \l   )  \label{psilk}
\ee
where the matrices $A, U$ have the form
\bea
&\&A(\l;x,\vec \l  ) = i\s_+ -i  \le(\l+\frac x 2 - \frac {\d a^{(n)}(x;\vec \l )}{\d x}\ri )\s_- + \sum_{j=1}^{n} \frac {A_j(x;\vec \l )}{\l-\l_j},
\\
&\& U(\l;x,\vec \l  )  =i\s_+  - i \le(\l -2 \frac{ \d a^{(n)}(x;\vec \l )}{\d x}\ri) \s_-.
\eea
The isomonodromic approach of \cite{JMU1, JMU2} proceeds as follows; one imposes  the {\it compatibility} of the equations \eqref{psil} \eqref{psix} \eqref{psilk}, namely, that there exists a {\it simultaneous} solution $\Psi_n(\l;x,\vec \l)$ of them. This requirement implies differential equations that determine the dependence on $x,\l_1,\dots, \l_n$ of the matrices $A, U, A_k$ appearing in the equations. The ensuing equations are usually referred to as ``zero curvature equations'' and take the following form
\be
\label{zeroc}
\pa_x A - \pa_\l U + [A,U] \equiv 0\ ,\qquad 
\frac{\pa_{\l_k} A_j}{\l-\l_j} - \frac{\pa_{\l_j}A_k}{\l-\l_j}  + \le[\frac{A_j}{\l-\l_j} , \frac {A_k} {\l-\l_k}\ri]\equiv 0 \nonumber \\
\pa_\l \frac {A_k}{\l-\l_k} - \pa_{\l_k} A + \le[\frac{A_k}{\l-\l_k}, A\ri] \equiv 0 \ ,\qquad
\frac{\pa_x A_k}{\l-\l_k} - \pa_{\l_k} U + \le[\frac{A_k}{\l-\l_k}, U\ri] \equiv 0.
\ee 
Viceversa, for any  collection of matrices $A,U, A_k$ satisfying \eqref{zeroc} there exists a joint solution $\Psi_n$ of equations (\ref{psil}, \ref{psix}, \ref{psilk}). 
Since the dependence on  $\l$ of $A$ is rational, the fundamental solution $\Psi_n$ of \eqref{psil} (normalized in some way that is not essential to specify now) is not necessarily single--valued: the analytic continuation of $\Psi_n$  along a  non-contractible contour $\gamma$  in $\C \setminus \{\l_1,\dots,\l_n\}$ yields a new matrix that solves the same ODE and hence it is related as $\Psi_n\mapsto \Psi_n  M_\gamma$. The matrix $M_\gamma$ depends only on the homotopy class and is called "monodromy matrix" associated to $\gamma$:  the collection of these matrices, for all homotopy classes, provides an (anti)-representation of the fundamental group of $\C \setminus \{\l_1,\dots,\l_n\}$.
In addition to these matrices one needs to compute the matrix Stokes' multipliers
(we refer to the introduction of \cite{JMU1} for a recall of this notion for the interested reader) and the collection of monodromy matrices and Stokes' multiplier is the ``generalized'' monodromy data: it is important to remind that these monodromy data are {\it independent} of $x,\l_1, \dots, \l_n$ precisely as a consequence of \eqref{zeroc}, so that they should be regarded as integrals of the motions.

In \cite {JMU1} the notion of isomonodromic tau function was then defined as follows; for any solution of \eqref{zeroc} (and associated $\Psi$--function) we can define the ``isomonodromic tau function'' $\tau_n(x;\vec \l)$ by means of 
\be
\label{JapTau}
\pa_{\l_k} \ln \tau_n (x;\vec \l ) &\&= \res{\l=\l_k} \Tr {A^2 \d \l}\ ;\qquad  
\pa_{x} \ln \tau_n (x;\vec \l ) = a^{(n)} (x; \vec \l  ).
\ee
The results of \cite{JMU1}  showed  (in a much more general setting) that
the equations \eqref{JapTau} form a compatible set of equations {\it provided that} the equations \eqref{zeroc} hold,   and hence they can be integrated to define $\tau_n(x;\vec \l)$ (which is, however, defined only up to multiplication by a scalar independent of $x, \vec \l$). The $\tau$ function depends parametrically on the generalized monodromy data (Stokes matrices and monodromy matrices) which replace the initial value conditions: the case that shall be of interest for us is when the monodromy representation is {\it trivial} and there is only the Stokes' phenomenon at $\l=\infty$. 

\subsection{Results}
At this point we can advertise the gist of our first result in the form of the following Theorem.
\bt
\label{main2}
Let $\tau_n(x;\vec \l  )$ be the isomonodromic tau function for the isomonodromic system (\ref{psil}, \ref{psix}, \ref{psilk}). Then the Kontsevich integral \eqref{ZnK} is equal to 
\be
Z_{n}(x;Y ) = {\rm e}^{\frac {x^3}{12}} \tau_n(x; \vec \l  ).
\ee
after the identification $y_k = \sqrt{\lambda_k},\, k=1,\ldots,n$.
\et

The formulation of the result in terms of isomonodromic deformation may be more widely recognizable by the readership, but it is not the way we want to set up its proof; 
the keen reader may also observe that the isomonodromic problem that we have indicated is still largely ambiguous because we did not, for example, specify the precise generalized monodromy data. 
Moreover, the isomonodromic formulation makes it hard to analyze the situation when the size of the matrix  integral in \eqref{ZnK} (the number of poles in \eqref{psil}) tends to infinity, which is our second main motivation to be discussed later on.

To remove all these ambiguities we will now  reformulate the isomonodromic system  (\ref{psil}, \ref{psix}, \ref{psilk}) directly in terms of a suitable Riemann--Hilbert problem, thus displaying explicitly  its monodromy data.
This reformulation allows to handle rigorously a limit as $n\to \infty$.
For technical reasons that should become clear later on, we shall formulate a slightly more general situation where the set $\l$ of $n$ points is partitioned in two  $ (\vec \lambda,\vec \mu) = (\lambda_1,\ldots,\lambda_{n_1},\mu_1,\ldots,\mu_{n_2})$ ($n=n_1+n_2$). 
Associated to this data we define the function
\be
\label{ddefinition}
	{\mathbf d}_n (\l)  := \prod_{j=1}^{n_1} \frac {\sqrt{\mu_j} + \sqrt{\l}}{\sqrt{\mu_j} - \sqrt{\l}} 
\prod_{j=1}^{n_2} \frac {\sqrt{\l_j} - \sqrt{\l}}{\sqrt{\l_j} + \sqrt{\l}}.
\ee

\begin{problem}
\label{RHPgamma}
Let $\Sigma$ be the union of oriented rays shown in Fig. \ref{jumpM}. 
Find a $2\times 2$  matrix valued analytic function $\G_{n} = \Gamma_{n}(\l;  \vec \l , \vec \mu )$  
such that:
\begin{enumerate}

\item[--] $\Gamma_{n}$ is locally bounded everywhere in  $\C$, 
 and analytic 
in  $\C \setminus \Sigma$.
\item[--] It admits continuous boundary values $\G_{n,\pm}$ on each ray and they satisfy the {\it jump conditions}
\bea
\G_{n}(\l)_+ &\& = \G_{n}(\l)_-  
M_{n}(\l)\ ,\ \ \ \ \l\in \Sigma,
\eea
where the matrix $M_{n}$ is piecewise defined by 
\bea 
M_{n} (\l) = \le\{\begin{array}{lc}
\1 + {\mathbf d}_n (\l){\rm e}^{- \frac 43 \l^\frac 32 - 2 x  \l^\frac 1 2   }\s_+  &  \l \in \varpi_0:=  {\rm e}^{i\theta_0}\R_+
\\[8pt]
\1 + \frac 1{{\mathbf d}_n (\l)}{\rm e}^{ \frac 43 \l^\frac 32+ 2 x  \l^\frac 1 2 }\s_-    & \l \in \varpi_\pm := {\rm e}^{i\theta_{\pm}}\R_+
 \\[8pt]
i\s_2  & \l\in \R_-.
\end{array}\ri. 
\label{Jump}
\eea
\item[--] Near $\l=\infty$, in each sector, it satisfies the following asymptotic expansion
\bea
\G_{n}(\l) &\& = \l^{-\frac {\s_3}4} \frac {\1 + i \s_1}{\sqrt{2}} \le(\1 + \frac{a^{(n)}(x;\vec \l, \vec \mu)}{\sqrt{\l}} \s_3 +\mathcal O(\l^{-1})\ri).
\label{Gninfty}
\eea
\end{enumerate}
\end{problem}

It is implied that the rays can be slightly deformed with respect to Fig. \ref{jumpM} so that 
none of the  poles of $\mathbf d_n (\l)$ lie on $\varpi_0$ and none of the poles of $\mathbf d^{-1}_n(\l)$ lie on $\varpi_{\pm}$. 
As a matter of fact the problem can be posed on arbitrary (non-intersecting) contours issuing from the origin and extending to infinity as long as the asymptotic directions at infinity are the ones indicated.

\br[Gauge arbitrariness]
\label{remgauge}
The asymptotic condition \eqref{Gninfty} implies a gauge fixing; indeed  we could multiply $\G_n$ on the left by a {\it constant} matrix  of the form $\1 + c \s_-$,  and this would not change the jump conditions. However that coefficient matrix of $\l^{-\frac 1 2}$ in the expansion \eqref{Gninfty} would be changed by the addition of a  term proportional to $\s_1$. 
In other words the requirement that the $\mathcal O(\l^{-\frac 1 2})$ term is proportional to $\s_3$ is part of the normalization condition at infinity (otherwise there would be a one--parameter family of solutions).
It is not hard to prove that the solution, if it exists, is unique under this normalization.
\er

We now explain how the Riemann--Hilbert problem \ref{RHPgamma} provides the precise (generalized) monodromy  data for the isomonodromic approach.   
The matrix 
\be
\label{defPsi}
\Psi_n &\&= \Psi_n(\l;x,\vec \l, \vec \mu ) := \Gamma_n (\l) {\rm e}^{ -\vartheta(\l;x) \s_3}
D^{-1}(\l),\qquad  \vartheta(\l;x) :=  \le( \frac 2 3 \l^{\frac 3 2 } + x \sqrt{\l}\ri)\\
 D(\l) = D(\l;\vec \l, \vec \mu)&\& := \le[
\begin{array}{cc}
\ds \prod_{j=1}^{n_2} (\sqrt{\l_j}+\sqrt{\l})\prod_{j=1}^{n_1} (\sqrt{\mu_j}-\sqrt{\l}) & 0 \\
0 &\ds  \prod_{j=1}^{n_2} (\sqrt{\l_j}- \sqrt{\l}) \prod_{j=1}^{n_1} (\sqrt{\mu_j}+\sqrt{\l})
\end{array}
\ri]\label{defD}  
\ee
satisfies a jump condition on $\Sigma$ with matrices independent of $\l, x, \vec \l, \vec \mu$. It then  follows  by standard arguments 
 that it satisfies an overdetermined system of PDEs  generalizing equations \eqref{psil}, \eqref{psix}, \eqref{psilk}, namely
\be
\frac {\pa}{\pa \l}\Psi_n(\l; x, \vec \l, \vec \mu)  &\& = A(\l;x, \vec \l, \vec \mu  ) \Psi _n(\l; x,\vec \l, \vec \mu) \label{psil2}\\
\frac {\pa}{\pa x}\Psi_n  (\l;x, \vec \l, \vec \mu  ) &\& = U(\l;x,\vec \l, \vec \mu ) \Psi_n (\l;x, \vec \l, \vec \mu ) \label{psix2}\\
\frac{\pa}{\pa \l_k}\Psi_n (\l;x, \vec \l, \vec \mu  ) &\&= -\frac{A_k (x ,\vec \l, \vec \mu )}{\l-\l_k}  \Psi_n(\l;x, \vec \l, \vec \mu  ), \; k=1,\ldots,n_1  \label{psilk2}\\
\frac{\pa}{\pa \mu_k}\Psi_n (\l;x, \vec \l, \vec \mu  ) &\&= -\frac{B_k (x ,\vec \l, \vec \mu )}{\l-\mu_k}  \Psi_n(\l;x, \vec \l, \vec \mu  ), \; k=1,\ldots,n_2, \label{psimuk}
\ee
where the matrices $A, U$ now have the form
\bea
&\&A(\l;x,\vec \l, \vec \mu ) = i\s_+ -i  \le(\l+\frac x 2 - \frac {\d a^{(n)}(x;\vec \l, \vec \mu)}{\d x}\ri )\s_- + \sum_{j=1}^{n_1} \frac {A_j(x;\vec \l, \vec \mu)}{\l-\l_j} + \sum_{j=1}^{n_2} \frac {B_j(x;\vec \l, \vec \mu)}{\l-\mu_j},
\label{AjBj}
\\
&\& U(\l;x,\vec \l, \vec \mu )  = i\s_+  - i \le(\l -2 \frac{ \d a^{(n)}(x;\vec \l, \vec \mu)}{\d x}\ri) \s_-.
\eea
and the function $a^{(n)}(x;\vec \l, \vec \mu)$ is defined (implicitly) above by the equation \eqref{Gninfty}.

These equations, together, represent a system of ``monodromy preserving'' deformation of the rational ODE \eqref{psil2}, in the sense of \cite{JMU1}. 

\br
\label{remspectrum}
While a general rational connection $\pa_\l - A$ with $A$  as in  \eqref{psil2} has nontrivial monodromy around the Fuchsian singularities of \eqref{psil}, our particular case corresponds to a situation  where the monodromy is trivial; more specifically, the residue matrices $A_j (x; \vec \l, \vec \mu )$ and $B_k (x;\vec \l, \vec \mu)$ have all eigenvalues $0$ and $\pm 1$,
since they were produced adding zeros and poles of order one in the first or the second column of the jump matrix $M_n$ in \eqref{Jump}. Thus the Fuchsian ODE is ``resonant'' \cite{Wasow}. The no-monodromy condition is a special constraint that determines the particular solution relevant to our problem.
\er
In this extended case the Jimbo--Miwa--Ueno definition of tau function translates to the following set of first order differential equations
\be
\label{JapTau2}
\pa_{\l_k} \ln \tau_n (x;\vec \l, \vec \mu) &\&= \res{\l=\l_k} \Tr {A^2 \d \l}\qquad
\pa_{\mu_k} \ln \tau_n (x;\vec \l, \vec \mu) = \res{\l=\mu_k} \Tr {A^2 \d \l}\nonumber \\
\pa_{x} \ln \tau_n (x;\vec \l, \vec \mu) &\& =  \res{\l=\infty}{ \Tr ( \Psi_n^{-1} \pa_\l \Psi_{n} \sqrt{ \l}\s_3)}  = a^{(n)} (x; \vec \l, \vec \mu )
\ee
generalizing straightforwardly the formul\ae\ \eqref{JapTau}.
Equations \eqref{JapTau2}, as it is customary in the Jimbo--Miwa--Ueno setting, determine the $\tau$ function up to a multiplicative factor that may depend on the monodromy data of the problem. To address this ambiguity, the definition was generalized in
 \cite {BertolaIsoTau} (with a correction in \cite{BertolaCorrection}) to one  that applies also to general Riemann--Hilbert problems:
\be
\label{Bertotau}
\pa \ln \tau_{_{JMU}} = \int_\Sigma \Tr \le(
\G_{n}^{-1} \G_{n}' \pa M_{n}  M_{n} ^{-1}
\ri) \frac {\d \l}{2i\pi},
\ee
  where $\Sigma = \R_- \cup \varpi_{0} \cup \varpi_+ \cup \varpi_-$. Since the jump on $\R_-$ in the Riemann--Hilbert problem is independent of parameters, the integration in \eqref{Bertotau} extends only on the three rays $\varpi_{0,\pm}$.\\
 
In this case the two definitions are completely equivalent but we will continue using the second one. 
Note that, however, the function $\tau$ is only defined up to multiplicative constants. 
In the cases where explicit integration of the above equation is possible, the integration constant will be tacitly set to zero, without further comment.

\paragraph{Extension of the Kontsevich matrix integral to arbitrary sectors.}
The right side of \eqref{ZnKY}, can be extended to an analytic function in the 
 left planes of the variables  because, up to the factor $\prod (y_j)^\frac 1 2$, the Airy functions are entire functions and the ratio in \eqref{ZnKY} is well defined on the ``diagonal'' sets $\{y_j=y_k,\ \  j,k=1,\dots,n\}$; however we now contend that we need to define it differently.
To explain the rationale we remind the reader  that the interpretation  of $Z_n(x;Y)$ as a generating function requires that it admits a {\it regular}\footnote{Here ``regular''  means that it is a  (formal) series in inverse powers of the $y_j's$, without exponential factors.} asymptotic expansion as $y_j\to \infty$.
Using the well-known asymptotic expansion of the Airy function ($\Ai$) in the sector  $|\arg\l| <\pi $ we see that 
\be
\nonumber 
\Ai(\l) = \frac{{\rm e}^{-\frac 2 3 \l^{\frac 32}}}{2\sqrt{\pi} \l^{\frac 1 4}}( 1 + \mathcal O(\l^{-\frac 3 2})) \  \ \Rightarrow\ \ 
{\rm e}^{\frac 23 y^3 + xy} \Ai(y^2 + x) =\le\{
\begin{array}{cc}
\ds 
 \frac{{\rm e}^{\frac 43 y^3 + 2xy}}{2\sqrt{\pi}\sqrt{y}}(\1 + \mathcal O(y^{-3})) & \arg y \in \le(\frac \pi 2, \frac {3\pi}2\ri) \\[10pt]
\ds  \frac{1}{2\sqrt{\pi}\sqrt{y}}(\1 + \mathcal O(y^{-3})) & \arg y \in \le(\frac{-\pi}2, \frac {\pi}2\ri) 
\end{array}\ri.
\ee
where we have used that $(y^2)^\frac 32 = -y^3$ if $\Re y\leq 0$ (and we use the principal roots).
Therefore \eqref{ZnKY}, as written, cannot possibly admit a regular asymptotic expansion if $y_j\to\infty$ in the sector $\Re y_j\leq 0$ for some $j$. 

The reader familiar with the Stokes' phenomenon of the Airy function will see that the way to recover  a regular expansion in the left half plane is to use either $\Ai(\omega^{\pm 1} y^2)$ instead.
To this end we introduce the following notation
\be
\AA_{\nu}(\l):= \Ai(\omega^\nu \l)\ ,\ \ \omega:= {\rm e}^{\frac {2i\pi}3}\ ,\ \ \nu =0,1,2.
\ee
The functions $\AA_\nu$ are solutions of the Airy equation  and satisfy $\AA_0 + \omega \AA_1 + \omega^2 \AA_2 \equiv 0$ and the functions $\sqrt{y} {\rm e}^{\frac 2 3 y^3 } \AA_\nu(y^2)$ admit a regular expansion  in inverse integer powers (without exponential terms) as $|y|\to\infty$ within the following sectors:
\be
\label{sectSS}
\begin{array}{ccc}
\mathcal S_0\!\! =\!\!\le\{\arg(y)\!\in\! \le(-\frac \pi 2 , \frac \pi 2\ri)\ri\};
&
 \mathcal S_1\!\!=\!\!\le\{\arg(y)\!\in\!\le(\frac \pi 6 , \frac {7\pi}6\ri)\ri\};
&
\mathcal S_2\!\! =\!\!  \le\{\arg(y)\!\in\!\le(\frac {5\pi}6 , \frac {11 \pi} 6\ri)\ri\}
\\
\raisebox{-0.45\height}{\begin{tikzpicture}[scale=0.4]
\draw circle[radius=1];
\draw [fill=gray!40!white] (0,-1) arc [radius =1, start angle= -90, end angle= 90] to (0,0) to (0,-1);
\end{tikzpicture}}
&
\raisebox{-0.45\height}{ \begin{tikzpicture}[scale=0.4]
\draw (0,0) circle[radius =1];
\draw [fill=gray!40!white] (0.86602540,0.5) arc [radius =1, start angle= 30, end angle= 210] to (0,0) to  (0.86602540,0.5) ;
\end{tikzpicture}}
&
 \raisebox{-0.45\height}{\begin{tikzpicture}[scale=0.4]
\draw (0,0) circle[radius =1];
\draw (0,0)--(0.5,-0.86602540);
\draw [fill=gray!40!white] (-0.86602540,0.5) arc [radius =1, start angle= 150, end angle= 330] to (0,0) to  (-0.86602540,0.5) ;
\end{tikzpicture}}.
\end{array}
\ee

\bd
\label{defWKcont}
For any partition of the set $\mathcal Y$ of the eigenvalues of $Y$ into three disjoint sets $\mathcal Y^{(s)},\ s=0,1,2$ of respective cardinality $n_0,n_1,n_2$ ($n= n_0+n_1+n_2$), we consider the following determinant which we call {\it  generalized  Kontsevich integral }
\be
\label{ZnKcont}
Z_n(x; \mathcal Y^{(0)}, \mathcal Y^{(1)}, \mathcal Y^{(2)} ) =
(-\omega)^{n_1 -n_2}
( 2\sqrt \pi)^{n}\frac{ {\rm e}^{\frac 23  \Tr Y^3  + x \Tr Y} \prod_{j=1}^{n} (y_j)^\frac 1 2 }{\prod_{j<k}(y_j-y_k)}\det \le[
\begin{array}{cc}
\ds \le[ \AA_0^{(k-1)} ( y_j^2 + x)\ri]_{y_j\in \mathcal Y^{(0)}\atop 1\leq k \leq n} \\
\hline 
\ds\le[ \AA_1^{(k-1)}( y_j^2 + x)\ri]_{y_{j}\in \mathcal Y^{(1)}\atop 1 \leq k \leq n}
\\
\hline 
\ds\le[ \AA_2^{(k-1)}( y_j^2 + x)\ri]_{y_{j}\in \mathcal Y^{(2)}\atop 1 \leq k \leq n}
\end{array}
\ri].
\ee
\ed
\noindent The generalized Kontsevich integrals \eqref{ZnKcont} reduce to  \eqref{ZnKY} if  $\mathcal Y^{(0)}= \mathcal Y$, $\mathcal Y^{(1)}=\emptyset= \mathcal Y^{(2)}$  and hence Theorem \ref{main2} is a special case of the theorem below.

\begin{shaded}
\bt\label{main2bis}\mbox{}\\
{\bf [1]} The  function $Z_n(x; \mathcal Y^{(0)}, \mathcal Y^{(1)}, \mathcal Y^{(2)} )$ \eqref{ZnKcont}   and  the isomonodromic tau function $\tau_n$ defined by \eqref{JapTau2} and   associated to the Riemann--Hilbert problem \ref{RHPgamma}
are related by 
\be Z_n(x; \mathcal Y^{(0)}, \mathcal Y^{(1)}, \mathcal Y^{(2)} ) = {\rm e}^{\frac {x^3}{12}} \tau_n(x; \mathcal Y^{(0)}, \mathcal Y^{(1)}, \mathcal Y^{(2)} ),\ee
with  the identification  $y_i = \sqrt{\lambda_i}\; \mathrm{if}\; \mathrm{Re}(y_i) >0$ and $y_j = -\sqrt{\mu_j}\; \mathrm{if}\; \mathrm{Re}(y_j) \leq 0$, all roots principal.

\noindent {\bf [2]} The expression  \eqref{ZnKcont}  admits  a regular asymptotic expansion  if the variables $y_j$'s tend to infinity provided that $
\mathcal Y^{(\nu)}  \subset \mathcal S_\nu$ with the sectors $\mathcal S_\nu$ defined in  \eqref{sectSS}

\noindent {\bf [3]} This asymptotic expansion is independent of the assignment of the variables to the different groups $\mathcal Y^{(\nu)}$ or $\mathcal Y^{(\wt \nu)}$  if they belong to the overlap of the sectors $\mathcal S_\nu \cap \mathcal S_{\wt \nu} $.
\et
\end{shaded}

The points {\bf [2]}, {\bf [3]}  of Thm. \ref{main2bis}  follow simply from the fact that ${\rm e}^{\frac 2 3 y^3 + xy} \Ai_{\nu,\wt \nu}(y^2 +x)$ have the {\it  same} regular asymptotic expansion  if $|y|\to\infty$ and $ y \in \mathcal S_\nu \cap \mathcal S_{\wt \nu} $. Indeed, 
the analysis of the asymptotic behaviour for $y\to \infty$ for the Airy function shows that  ${\rm e}^{\frac  2 3 y^3} \AA_\nu (y^2)$  admits the same  regular (nontrivial) expansion in integer inverse powers of $y$'s if and only if 
$y$ tends to infinity in the corresponding sectors $\mathcal S_\nu $  (see \cite{AbramowitzStegun}, 10.4.59).

In particular, if all $y_j$'s tend to infinity in  the right half-plane and we assign them all to $\mathcal S_0$, then we get \eqref{ZnK} and hence this expansion {\it is the formal expansion} that generates the intersection numbers of tautological classes as explained in \cite{Kontsevich:1992p30}.

Using the alternative but equivalent formula \eqref{Bertotau} we can restate the Theorem \ref{main2bis} in the form 
\bt
\label{main}
The Kontsevich integral  $Z_n(x;\mathcal Y^{(0)},\mathcal Y^{(1)},\mathcal Y^{(2)})$ in \eqref{ZnKcont} satisfies
\be
\pa \ln Z_n(x;\vec \l, \vec \mu )
&\&=\pa \frac { x^3}{12} + 2\int_{\varpi_0} \le(
 \G^{-1}_n(\l) \G'_n(\l)\ri)_{21}
\pa  {\mathbf d}_n  (\l) {\rm e}^{- \frac 43 \l^{\frac 32} - 2 x  \l^\frac 1 2}
 \frac {\d \l}{2i\pi}
 +\cr
&\& + 2\sum_{\pm}
 \int_{\varpi_\pm } \le(
 \G^{-1}_n(\l) \G'_n(\l)\ri)_{12} 
 \pa  {\mathbf d}_n^{-1} (\l) {\rm e}^{ \frac 43 \l^{\frac 32}  + 2 x  \l^\frac 1 2}
 \frac {\d \l}{2i\pi} 
\label{varZn}
\ee
where $\pa$ is the derivative with respect to any parameter $x,\{\vec \lambda,\vec  \mu\}$.
The relationship between the parameters $\{y_j\}$ and $\{\vec \l, \vec \mu\}$ is $y_i = \sqrt{\lambda_i}\; \mathrm{if}\; \mathrm{Re}(y_i) >0$ and $y_j = -\sqrt{\mu_j}\; \mathrm{if}\; \mathrm{Re}(y_j) \leq 0$.
\et
\noindent For the proof see Sec. \ref{proofmain}.

\paragraph{The limit $n\to\infty$: first Painlev\'e\ hierarchy.}
It was one of the main points of  Kontsevich's original work \cite{Kontsevich:1992p30}  that the integral  \eqref{ZnK} is formally a KdV tau function in the Miwa variables\footnote{The factor $2^{-\frac{2k + 1}3}$ stems from our normalization,  see Remark \ref{normalisation}.}
\be\label{Miwa}
T_k(Y):= -\frac {2^{-\frac{2k + 1}3}}{(2k+1)!!} \Tr Y^{-2k-1}.
\ee 
More precisely, in these variables, the function $U(x;T) := \frac{\partial^2}{\partial {T_0}^2}\log Z_n(x;Y)$ satisfies the KdV hierarchy with the normalization adopted in \cite{Witten91}.
This particular solution of the KdV hierarchy was known by physicists even before the formulation of Witten's conjecture and its proof by Kontsevich \cite{Kontsevich:1992p30}. In the physics literature, this is referred to as the {\it partition function of 2D topological gravity} (see the references in \cite{DijkgraafReview}). It can be  defined as the solution satisfying the initial value condition
$U(x,0) = x$. As originally discovered by Douglas \cite{Douglas} using the so--called {\it string equation} (see Section \ref{secP1N} below, keeping in mind that there the normalization is different from Witten's, see Remark \ref{normalisation}), the Witten--Kontsevich solution of the KdV hierarchy satisfy an infinite number of ODE's in $T_0$ (or $x$, which is the same) known as the Painlev\'e I hierarchy, and where the higher $T_i$'s play the role of parameters. Indeed this is very close to the procedure of Flaschka and Newell \cite{FlaschkaNewell} who deduce the Painlev\'e II hierarchy as a self--similar reduction of the modified KdV one. To see briefly how it works, recall that, in the Witten's normalization, the equations of the KdV hierarchy are written as 
\be\label{KdVWitten}
	\frac{\partial U}{\partial T_i}  = \frac{\partial R_{i+1}}{\partial T_0},\; i\geq 0, 
\ee
where the $R_i$ are differential polynomials  in $U(T_0)$ defined by the recursion
\be\label{LenardWitten}
	\frac{\partial R_{k+1}}{\partial T_0} = \frac{1}{2k + 1}\left(\frac{\partial U}{\partial T_0} + 2 U \frac{\partial}{\partial T_0} + \frac{1}4 \frac{\partial^3}{\partial T_0^3}\right)R_k; \quad R_1(U) = U.
\ee
Besides these equations, the function  $F(x;T) := \log Z_n(x;Y)$ satisfies also the first Virasoro constraint (which can be deduced as a consequence of the fact that $Z_n(x;Y)$ is a matrix integral)
\be\label{SEWitten}
	\frac{\partial F}{\partial T_0} = \frac{T_0^2}{2} + \sum_{i = 0}^\infty T_{i + 1}\frac{\partial F}{\partial T_i}.
\ee
Differentiating once \eqref{SEWitten} with respect to  $T_0$ and substituting the integrated version of \eqref{KdVWitten} (integration constants are seen to be equal to zero) one obtains the set of ODE's in $T_0$, depending on the parameters $\{T_i\}$,
\be\label{PIWitten}
	T_0 + (T_1 - 1)U + \sum_{i \geq 1} T_{i + 1} R_{i + 1} = 0.
\ee
More precisely the $N$-th member of the Painlev\'e I hierarchy is obtained by putting $T_j = 0$ when $j\geq N+1$  and it is an ODE in $x=T_0$ depending parametrically on $T_1, \dots, T_N$ (more details are recalled in Section \ref{secP1N}).
Making sense of the formal statement ``the Kontsevich matrix model $Z_n(x;Y)$ satisfies the Painlev\'e I hierarchy'', requires a limit $n\to\infty$  because, for fixed $n$,  the variables $T_j$ are not even independent. 
This leaves open the question as to what kind of convergence we should expect. Also, $Z_n(x;Y)$ is usually treated as a formal series, while it would be interesting to analyze the analytic properties of these solutions of the hierarchy.
Thus the issue becomes:
\begin{question}
\label{Q1}
\noindent How to choose a sequence $\mathcal Y = \mathcal Y_n$ of variables  $\mathcal Y_n = \le\{y_1^{(n)},\dots, y_n^{(n)}\ri\} $  and an appropriate partition into three subsets in such a  way that generalized Kontsevich integral \eqref{ZnKcont} converges to a tau function of the $N$-th member of the first Painlev\'e\ hierarchy?  Which particular solution does it converge  to  and in which domain of the parameters? 
\end{question}
\noindent  To address the question, let $r\in \N$ and $P_r$ be the $r$-th Pad\'e\ approximant to ${\rm e}^{-z}$  (see \eqref{Padexpn}), which is a polynomial of degree $r$.  
Denote by $a_1,\dots, a_r$ its zeroes. It is known \cite{SaffVarga78} that they are all in the region  $\Re z>0$ (see Fig. \ref{plotzeros}).
Fix $N\in \N$ and set 
\be
\label{zerosP1N}
\mathcal Y = \{y_1,\dots, y_{n}\} =\le \{y:\ \ \ P_r( 2t  y^{2N+1})= 0 \ri\}\ ,\ \ \  n = r(2N+1).
\ee
The set $\mathcal Y$ is naturally  partitioned into subsets of cardinality $r$ as follows (see Fig. \ref{EVENNfig} and Fig. \ref{ODDNfig})
\be
\label{Ykappa}
\mathcal Y_\kappa:= \le\{ {\rm e}^{\frac {i\pi \kappa}{2N+1} }\le(\frac{a_j}{2t}\ri) ^{\frac 1{2N+1}} ,\ \ 1\leq j \leq r\ri\} ;
\ \ \arg \mathcal Y_k \subset \le( -\frac \pi {4N+2}, \frac \pi{4N+2} \ri) - \frac{\arg t}{2N+1}
\\
\kappa = -N, \dots,N-1,  N ,
\ee
 where $\arg (\mathcal Y_k)$,   denotes the set of the arguments of the elements in $\mathcal Y_k$.

\noindent We then address Question \ref{Q1} for the subsequence $n = r(2N+1)$  and let $r\to \infty$ while keeping $N$ fixed.
  For a brief review of the  first Painlev\'e\ hierarchy, its Riemann--Hilbert formulation and associated tau function we refer the reader  to Section \ref{secP1N}.

\begin{figure}
\begin{minipage}{0.47\textwidth}
\begin{center}
\begin{tikzpicture}[scale=1.5]
\foreach \x in {-4.75,-3.75,-2.75,-1.75,-0.75,0.25, 1.25, 2.25, 3.25}
{
\draw[dashed]  ({2*\x*360/14}:0.4) to ({2*\x*360/14}:1.7);
\draw[dashed]  ({2*\x*360/14 + 360/14}:0.4) to ({2*\x*360/14+ 360/14}:1.7);
};
\fill [line width=0pt, fill= white!17!black!70!green!65, fill opacity = 0.3]  (0,0) to (30:1.6)  arc (30:210:1.6);
\fill [line width=0pt, fill= white!17!black!70!red!65, fill opacity = 0.3]  (0,0) to (-30:1.5)  arc (-30:-210:1.5);
\fill [line width=0pt, fill= white!17!black!70!blue!65, fill opacity = 0.3]  (0,0) to (-90:1.7)  arc (-90:90:1.7);
\draw [dashed, <->]   ({360/14*2.5}:1)  arc ({360/14*2.5}:{-360/14*2.5}:1)  ;
\node at (360/14*1:1.2) {$\mathcal Y^{(0)}$};
\draw [dashed, <->]   ({360/14*1.5}:1.6)  arc ({360/14*1.5}:{360/14*2.5}:1.6)  ;
\node [right] at (360/14*2:1.6) {$\mathcal Y_\kappa$};
\draw [dashed, <->]   ({360/14*3.5}:1)  arc ({360/14*3.5}:{360/14*6.5}:1)  ;
\node at (360/14*5:1.2) {$\mathcal Y^{(1)}$};
\draw [dashed, <->]   ({-360/14*3.5}:1)  arc ({-360/14*3.5}:{-360/14*6.5}:1)  ;
\node at (-360/14*5:1.2) {$\mathcal Y^{(2)}$};
\draw (-2,0)--(2,0);
\foreach \r in {-3,-2,-1,0,1,2,3}
{
\begin{scope}[rotate= \r*360/7]
\foreach \x in 
{(1.5736301664, -.2949198865), (1.5736301664, .2949198865), (1.5621330212, -.2410187845), (1.5621330212, .2410187845), (1.5543664266, -.1951001010), (1.5543664266, .1951001010), (1.5487899680, -.1530420157), (1.5487899680, .1530420157), (1.5448060416, -.1132389050), (1.5448060416, .1132389050), (1.5421152933, -0.748219047e-1), (1.5421152933, 0.748219047e-1), (1.5405554649, -0.372206448e-1), (1.5405554649, 0.372206448e-1), (1.5400440083, 0.) }
{
\draw [fill]\x circle[radius=0.3pt];
};
\end{scope}};
\end{tikzpicture}
\end{center}
\captionof{figure}{Example with $N=3$: the assignment of the $\mathcal Y_\kappa$'s to the disjoint subsets $\mathcal Y^{(0,1,2)}$ indicated in the figure corresponds to $k_+ =1, k_0=0, k_-=-1$.}
\label{ODDNfig}
\end{minipage}
\hfill
\begin{minipage}{0.47\textwidth}
\begin{center}
\begin{tikzpicture}[scale=1.5]
\foreach \x in {-4.75,-3.75,-2.75,-1.75,-0.75,0.25, 1.25, 2.25, 3.25}
{
\draw[dashed]  ({2*\x*360/18}:0.4) to ({2*\x*360/18}:1.7);
\draw[dashed]  ({2*\x*360/18 + 360/18}:0.4) to ({2*\x*360/18+ 360/18}:1.7);
};

\fill [line width=0pt, fill= white!17!black!70!green!65, fill opacity = 0.3]  (0,0) to (30:1.6)  arc (30:210:1.6);
\fill [line width=0pt, fill= white!17!black!70!red!65, fill opacity = 0.3]  (0,0) to (-30:1.5)  arc (-30:-210:1.5);
\fill [line width=0pt, fill= white!17!black!70!blue!65, fill opacity = 0.3]  (0,0) to (-90:1.7)  arc (-90:90:1.7);
\draw [dashed, <->]   ({360/18*2.5}:1)  arc ({360/18*2.5}:{-360/18*2.5}:1)  ;
\node at (360/18*1:1.2) {$\mathcal Y^{(0)}$};
\draw [dashed, <->]   ({360/18*1.5}:1.6)  arc ({360/18*1.5}:{360/18*2.5}:1.6)  ;
\node [right] at (360/18*2:1.6) {$\mathcal Y_\kappa$};
\draw [dashed, <->]   ({360/18*3.5}:1)  arc ({360/18*3.5}:{360/18*8.5}:1)  ;
\node at (360/18*6.5:1.2) {$\mathcal Y^{(1)}$};
\draw [dashed, <->]   ({-360/18*3.5}:1)  arc ({-360/18*3.5}:{-360/18*8.5}:1)  ;
\node at (-360/18*6.5:1.2) {$\mathcal Y^{(2)}$};

\draw (-2,0)--(2,0);

\foreach \r in {-4,-3,-2,-1,0,1,2,3,4}
{
\begin{scope}[rotate= \r*360/9]
\foreach \x in 
{(1.4270937301, -.2070717788) , (1.4270937301, .2070717788) , (1.4176118351, -.1695874370) , (1.4176118351, .1695874370) , (1.4111552221, -.1374793587) , (1.4111552221, .1374793587) , (1.4064926635, -.1079578715) , (1.4064926635, .1079578715) , (1.4031475642, -0.799417690e-1) , (1.4031475642, 0.799417690e-1) , (1.4008814933, -0.528486994e-1) , (1.4008814933, 0.528486994e-1) , (1.3995653072, -0.262979487e-1) , (1.3995653072, 0.262979487e-1) , (1.3991333293, 0.) }
{
\draw [fill]\x circle[radius=0.3pt];
};
\end{scope}};
\end{tikzpicture}
\end{center}
\captionof{figure}{Example with $N=4$; the assignment of the $\mathcal Y_\kappa$'s to the disjoint subsets $\mathcal Y^{(0,1,2)}$ indicated in the figure corresponds to $k_+ =1, k_0=0, k_-=-1$. }
\label{EVENNfig}
\end{minipage}
\end{figure}

\begin{shaded}
\bt
\label{ThmP1N}
Fix $N\in \N$. Partition the set of indices $\kappa$  (with $-N\leq \kappa\leq N$  modulo $2N+1$) into  three consecutive  
disjoint groups   such that the corresponding subsets $\mathcal Y_\kappa$'s \eqref{Ykappa} belong  to the sectors $\mathcal S_{0,1,2}$ \eqref{sectSS}; let    $\mathcal Y^{(0,1,2)}$ be their respective  unions, correspondingly. 
Define   $k_+\geq k_0 \geq k_-,\ k_+> k_-$ as follows
\begin{itemize}
\item $k_0= $ (number of $\mathcal Y_\kappa$'s in the second quadrant that we assign  to $\mathcal Y^{(2)}$)  - (number of   $\mathcal Y_\kappa$'s in the third  quadrant that we assign to $\mathcal Y^{(1)}$)
\item $k_- = -\le\lfloor \frac {N}2 \ri \rfloor$ +  ( number of  $\mathcal Y_\kappa$'s in the first quadrant that we assign to $\mathcal Y^{(1)}$ );
\item $k_+ = \le\lfloor \frac {N}2 \ri \rfloor$ -  ( number of $\mathcal Y_\kappa$'s  in the fourth quadrant that we assign to $\mathcal Y^{(2)}$ );
\end{itemize}

Then  
 \begin{enumerate}
\item the formula \eqref{ZnKcont} 
 converges, with rate of convergence $\mathcal O(n^{-\infty})$  as $n\to \infty$, to the tau function $\tau(x;t)$ of the special tronqu\'ee solution of the $N$-th member of the $PI$ hierarchy defined  by the formula \eqref{defTauP1N} in terms of the solution of  the RHP \ref{RHPP1Nspecial} with the chosen $(k_+,k_0,k_-)$. Then the function   $ u(x,t) := 2 \pa_x^2 \ln \tau(x;t)$ satisfies the nonlinear ODE
\be
\label{P1N}
( {2N+1})  t \L_{N}[u(x;t )] +  u(x;t) + x = 0 \ .
\ee
\item All these particular solutions $u(x;t)$  have no poles for $|t|$ sufficiently small within an open    sector  of width at least $ \pi $ that contains $\arg (t)=0$.
Within the common sector where they have no poles they differ, as $|t|\to 0$, by  $\mathcal O(|t|^{\infty})$ terms . If $k_0=0$, then the width of this sector is at least $\pi$ on either sides of $\arg t=0$.  
\item The limit at $t=0$, from within this common sector, of the  derivatives of arbitrary order equal those of the  formal topological solution.
\end{enumerate}
\et
\end{shaded}

\noindent For the proof see Sec. \ref{proofP1N}. Here $\L_N[u]$ is the Lenard differential polynomial in $u(x)$ whose definition will be reviewed below in \eqref{Lenard}.

The idea behind the  binning of the groups $\mathcal Y_\kappa$ into the three disjoint subsets $\mathcal Y^{(0,1,2)}$ of Theorem \ref{main2bis} and \ref{ThmP1N}  is as follows; referring to the Figures \ref{ODDNfig}, \ref{EVENNfig}  we see that some groups can only be assigned to one $\mathcal Y^{(\nu)}$ because they belong to only one $\mathcal S_\nu$, while others fall in the intersection between two different sectors $\mathcal S_{0,1,2}$ and can be assigned to either. The  different choices are reflected in the choices of the parameters $k_+,k_0,k_-$ that characterize the particular tronqu\'ee solution in the Riemann--Hilbert problem \ref{RHPP1Nspecial}. The Theorem \ref{ThmP1N}  is therefore a first foray in the study of the nonlinear Stokes' phenomenon for the Witten--Kontsevich tau function.

\br
From the recurrence relation \eqref{Lenard} we can see that the Lenard polynomials $ \L_N[u]$ are homogeneous of degree $2N$ under the rescaling $U(X)   = \alpha^{2} u(\alpha^{-1} X)$. 
Setting $U(X,t):= t^{\frac {2}{2N+1}}  u(t^{-\frac 1 {2N+1}} X, t)$, we obtain the equation 
\be
( {2N+1})   \L_{N}[U(X;t )] + t^{-\frac {3}{2N+1}} U(X;t) + X = 0 
\ee
and this shows that $u(x,t)$ is single-valued on the Riemann surface of $t^{\frac 1{2N+1}}$. This explains how it is possible to have no poles in a sector of amplitude $2\pi$ or even bigger.
\er
\begin{example}
For $N=2,3$ the equation \eqref{P1N} reads
\be
N=2; && \frac 5 {8} t \le(u'' + 3u^2
\ri) +  u + x = 0
\\
N=3; &&\frac 7 {32} t\le(
u^{(4)} + 10 u u''  + 5 (u')^2 + 10 u^3
\ri) +    u+x=0\ . 
\ee
The  case $N=2$ above is, up to the map $u(x) = \le(\frac {8} {5 t }\ri)^{\frac 2 5} U(X)- \frac {4}{15 t},\ 
x =  -\le(\frac{t}8\ri)^\frac 1 5  X- \frac { 2}{15 t}$ 
the standard first Painlev\'e\  1 equation $U'' + 3U^2 = X$; in this case the particular solution is precisely a tritronqu\'ee solution \cite{ItsKapaevFokasBook}. 
\end{example}

In Thm. \ref{ThmP1N} we restricted ourselves to the subsequence $n=r(2N+1)$ only because of the way we constructed the rational approximation of ${\rm e}^{2t \l^{\frac {2N+1}2}}$; to extend the statement to the whole sequence  one would have to consider the Pad\'e\ approximants to ${\rm e}^{2t z^{2N+1}}$ directly (of which, the polynomials $P_r(2t z^{2N+1})$ are a subsequence). 
More generally, the full fledged member of the Painlev\'e\ hierarchy as in \eqref{PIWitten} would require the  analog of the estimate \eqref{EstimPade} for the location of the zeroes and the estimate of the remainder term  of the general exponential ${\rm e}^{\sum_{j=1}^{N} t_{2j+1} z^{2j+1}}$. 
We regard this issue as a technical one; we expect that the general phenomenon will be the same we observe in this restricted case.

\paragraph{Acknowledgements.} 
The research of M.B  was supported in part by the Natural Sciences and Engineering Research Council of Canada grant
RGPIN/261229--2011 and by the FQRNT grant "Matrices Al\'eatoires, Processus Stochastiques et Syst\`emes Int\'egrables" (2013--PR--166790).
The research of M.C. was partially supported by a project ``Nouvelle \'equipe'' funded by the region Pays de la Loire. M.C. thanks the Centre de Recherches Math\'ematiques (CRM)  in Montr\'eal and the International School of Advanced Studies (SISSA) in Trieste for their hospitality during the preparation of this work.

\section{The Riemann--Hilbert problem for the first Painlev\'e\ hierarchy and associated $\tau$ function}
\label{secP1N}
In order to discuss the various solutions of the first Painlev\'e\ hierarchy, we need to review the relevant Riemann--Hilbert problem (\cite{ItsKrasovskyClaeys} , page 365).
The Riemann--Hilbert problem of the  $N$-th member of the first Painlev\'e\ hierarchy is constructed as follows;
define the {\it phase function}
\be
\label{phaseP1N}
\vartheta(\l) := t_{_{2N+1}} \l^{\frac {2N+1}2} + \sum_{j=0}^{N-1} t_{2j+1} \l^{\frac {2j+1}2} \ ,\ \ \ t_1:= x.
\ee
and let $\varpi_\nu$ be the rays $\arg( \l ) -\frac {2\arg (t)}{2N+1} = \frac{2\pi \nu}{2N+1}$, $-N\leq \nu \leq N$.  
\begin{problem}[First Painlev\'e\ hierarchy]
\label{RHPP1N}
Find a $2\times 2$ matrix $\G(\l)$, locally bounded everywhere in $\C$, 
   analytic away from the rays $\varpi_\nu$  (oriented towards infinity)  and $\R_-$ (oriented towards the origin) and such that 
\begin{itemize}
\item [--] it admits non-tangential boundary values at the points of the rays and they satisfy
\be
\G_+(\l) = \G_- {\rm e}^{-\vartheta(\l)\s_3} S_{\nu} {\rm e}^{\vartheta(\l)\s_3}\ ,\ \ \ \  \l \in\varpi_\nu\ ,\qquad 
\G_+(\l) = \G_-(\l) i\s_2,\ \ \ \ \l\in \R_-
\ee
with
\be
S_\nu=  \le\{\begin{array}{ll}
S_{2j} =  \le[\begin{array}{cc}
1 & s_{2j}\\
0&1
\end{array}\ri] & \nu = 2j\\
S_{2j+1}=\le[\begin{array}{cc}
1 & 0\\
s_{2j+1}&1
\end{array}\ri]
&
\nu =2j+1.
\end{array}
\ri.\ ,\qquad 
\nu =-N  \dots, N,
\ee 
and such that the $2N+2$ parameters $s_{-N}, \dots s_{N}$ are subject only to the {\it no monodromy condition }
\be
\label{nomonodromy}
S_{-N} \cdots S_0\cdots S_N = i\s_2.
\ee
\item [--]
Near $\l=\infty$ the solution has the same sectorial asymptotic expansion  in each sector, normalized by 
\be
\G(\l;\t) = \l^{-\frac {\s_3}4} \frac {\1 + i\s_1}{\sqrt{2}} \le(\1 + a(\t)\frac{\s_3}{\sqrt{\l}} + \mathcal O(\l^{-1}) \ri).
\ee
\end{itemize}
\end{problem}
\noindent
The connection with the equations of the hierarchy arises as follows. 
The tau function of a solution corresponding to the above data is defined in \cite{JMU1} as 
\be
\label{defTauP1N}
\pa_{t_j} \ln \tau_{P1_N}(\t) = -  \res{\l=\infty} \Tr \le(\l ^{\frac {2j+1}2} \Gamma^{-1}(\l;\t) \Gamma'(\l;\t) \s_3  \ri) \d \l,
\ee
where the residue is to be intended as a formal one (the formal series in the residue turns out to have only integer powers of $\l$ and the residue is the coefficient of the power $\l^{-1}$ of the expression in the bracket). 
The function 
\be
u(x,t_3,t_5,\dots, t_{2N+1}) := 2\pa_x a (x,t_3,t_5,\dots, t_{2N+1})  = 2 \pa_x ^2 \ln \tau_{P1_N}((x,t_3,t_5,\dots, t_{2N+1}) \label{222}
\ee
satisfies the following ODE in $x=t_1$, depending parametrically on $t_3,\dots, t_{2N+1}$:
\be
\sum_{k=1}^{N} (2k+1) t_{2k + 1} \L_k[u] + x=0\ . 
\label{P1Ngeneral}
\ee
Here $\L_k[u]$ are the Lenard-Magri differential polynomials defined \cite{DickeyBook}
by the recursion relations:  
\begin{equation}\label{Lenard}
	\frac{\pa}{\pa x} \L_{n+1}[u] = \le(\frac 1 4 \frac{\pa^3 }{\pa x^3 }+ u(x)\frac{\pa}{\pa x} + \frac 1 2 u_x(x)\ri) \L_n[u],\ \ \ \L_0[u]= 1,\ \ \ \L_n[0]=0
\end{equation}
The Stokes' parameters $\vec s = (s_{-N}, \dots, s_N)$ (subject to   \eqref{nomonodromy}) parametrize the solution space of 
\eqref{P1Ngeneral}.

In addition to the ODE \eqref{P1Ngeneral} above, $u$ satisfies also  
\begin{equation}\label{KdV}	\frac{\pa u}{\pa t_{2j+1}} =2 \frac \pa{\pa x} \L_{j+1}[u],\ \ \ j \leq N;\ \ \ u=u(\t ),\ \ \t =(t_1,t_3,t_5,\ldots).
\end{equation}

\br\label{normalisation}
The equations \eqref{P1Ngeneral}, \eqref{Lenard} and \eqref{KdV}, up to a rescaling and a shift of $T_1 = t_3$, correspond to \eqref{PIWitten}, \eqref{LenardWitten}, \eqref{KdVWitten}.
The source of the difference comes  from  the normalization we have used for the matrix integration variable in \eqref{ZnK}; indeed Kontsevich writes the integrand as $\exp \le[\frac i 6 \Tr X^3 - \frac 1 2 \Tr X\Lambda X\ri]$, from which we conclude that the relationship between our $M, Y$ and his $X, \Lambda$ is 
\be
M = 2^{-\frac 13} X,\ \  Y = 2^{-\frac 1 3} \Lambda. 
\ee
This translates to the following scaling relationship for the times (comparing our phase function \eqref{phaseP1N} with the phase function in \cite{BertolaDiDubrovin}, eq. (1.9), which yields the correct normalizations) 
\be
t_{2j+1} = -\frac{ 2^{\frac {2j+1}3} (T_j - \delta_{j,1} ) }{(2j+1)!!}. 
\ee
In particular our $t_1 = x$ corresponds to $-2^\frac 13 T_0$.
\er
\br
The equation \eqref{P1Ngeneral}  is the statement that the tau function of the solution to RHP \ref{RHPP1N} is the reduction of a Korteweg--de Vries (KdV) tau function satisfying the string equation
$[P,L] = 1,$
where $L := \frac{\partial^2}{\partial x^2} + 2u(x,t_1,\ldots,t_{k})$ is the Lax operator for the KdV hierarchy and $P := \sum_{k = 1}^N (2k + 1)t_{2k + 1} L^{\frac{2k + 1}2}_+$, see for instance \cite{MooreString}. The formulation of the Painlev\'e I hierarchy in terms of the string equation is originally due to Douglas \cite{Douglas}.
\er

\bx
The case $N=2$ corresponds to the first Painlev\'e\ equation and the special solution  is the famous {\it tri-tronqu\'ee} solution \cite{ItsKapaevFokasBook}. Also, for all even $N$ these are the solutions (conjecturally) relevant to the study of the ``higher order'' critical behavior of the largest eigenvalue in certain random matrix models \cite{ItsKrasovskyClaeys}.
\ex
\noindent We shall need also the formula for the higher derivatives of $\ln \tau_{P1_N}(\t)$; this formula is explained in \cite{BertolaDiDubrovin} in the more general context of the KdV hierarchy (of which the Painlev\'e\ I hierarchy is a reduction).
\be
\label{higherderivatives}
\frac {\pa^{k}}
{\pa t_{2j_1+1} \dots \pa t_{2j_k+1} }
\ln \tau_{P1_N}(\t)  = \prod_{j=1}^{k} \res{\l_\ell=\infty} \l_\ell^{j_\ell} F_k(\l_1,\dots,\l_k)
\ee
\be
 F_k(\l_1,\dots,\l_k)=-\frac{1}{k}\sum_{\rho\in S_{k}}  \frac{{\rm Tr} \le( \Theta(\l_{r_1})\cdots \Theta(\l_{\rho_k})\ri)}{
\prod_{j=1}^k(\l_{\rho_j}-\l_{\rho_{j+1}} )} -\delta_{k,2}\frac{\l_1+\l_2}{(\l_1-\l_2)^2}, \quad k\geq 2. \label{no-n}
\ee
\be
\Theta(\l) =\Theta(\l; \t) = \Gamma (\l;\t )\s_3\Gamma ^{-1}(\l;\t).
\ee
where $S_k$ is the permutation group of $k$ elements and in the formula we convene that $\rho_{k+1} \equiv \rho_1$.

\section{Kontsevich's integral as isomonodromic tau function}
This section contains the proof of the main theorems \ref{main2bis} and \ref{ThmP1N}, presented respectively in the subsections 3.3 and 3.4. 

\subsection{The bare solution: Airy RHP}
\label{sectionbare}
\begin{figure}
\begin{center}
\resizebox{0.42\textwidth}{!} {
\begin{tikzpicture}[scale=2.3]
\fill [line width=0pt, fill= white!7!black!80!red!80, fill opacity = 0.3]  (0,0) to (60:2) 
arc (60:300:2);
\draw (-2,0)--(2,0);
\node at (160:1) {II};
\node at (190:1) {III};
\node at (60:1) {I};
\node at (-60:1) {IV};
\draw [ line width = 1.5pt, postaction={decorate,decoration={markings,mark=at position 0.65 with {\arrow[line width=1.5pt]{>}}}}] (0,0) to node [pos =0.8, above,sloped] {$\le(\begin{array}{cc} 1&1\\ 0&1\end{array}\ri)$} (04:2);
\draw [ line width = 1.5pt, postaction={decorate,decoration={markings,mark=at position 0.65 with {\arrow[line width=1.5pt]{<}}}}] (0,0) to node [pos =0.8, above,sloped] {$\le(\begin{array}{cc} 1&0\\1&1\end{array}\ri)$} (140:2);
\draw [ line width = 1.5pt, postaction={decorate,decoration={markings,mark=at position 0.65 with {\arrow[line width=1.5pt]{<}}}}] (0,0) to node [pos =0.8, below,sloped] {$\le(\begin{array}{cc} 1&0\\1&1\end{array}\ri)$} (204:2);
\draw [ line width = 1.5pt, postaction={decorate,decoration={markings,mark=at position 0.65 with {\arrow[line width=1.5pt]{<}}}}] (0,0) to node [pos =0.8, above,sloped] {$\le(\begin{array}{cc} 0&1\\ -1&0\end{array}\ri)$} (180:2);
\end{tikzpicture}}\hfill
\resizebox{0.42\textwidth}{!} {
\begin{tikzpicture}[scale=2.3]
\fill [line width=0pt, fill= white!7!black!80!red!80, fill opacity = 0.3]  (0,0) to (60:2) 
arc (60:300:2);
\draw (-2,0)--(2,0);
\node at (160:1) {II};
\node at (190:1) {III};
\node at (60:1) {I};
\node at (-60:1) {IV};
\draw [ line width = 1.5pt, postaction={decorate,decoration={markings,mark=at position 0.65 with {\arrow[line width=1.5pt]{>}}}}] (0,0) to node [pos =0.6, above,sloped] {$\le(\begin{array}{cc} 1&{\rm e}^{-\frac 43 \l^\frac 32 - 2x \sqrt\l} \\  0&1\end{array}\ri)$} (04:2);
\draw [ line width = 1.5pt, postaction={decorate,decoration={markings,mark=at position 0.65 with {\arrow[line width=1.5pt]{<}}}}] (0,0) to node [pos =0.6, above,sloped] {$\le(\begin{array}{cc} 1&0\\ {\rm e}^{\frac 43 \l^\frac 32 +2x \sqrt\l} &1\end{array}\ri)$} (140:2);
\draw [ line width = 1.5pt, postaction={decorate,decoration={markings,mark=at position 0.65 with {\arrow[line width=1.5pt]{<}}}}] (0,0) to node [pos =0.6, below,sloped] {$\le(\begin{array}{cc} 1&0\\ {\rm e}^{\frac 43 \l^\frac 32 + 2x \sqrt\l} &1\end{array}\ri)$} (204:2);
\draw [ line width = 1.5pt, postaction={decorate,decoration={markings,mark=at position 0.65 with {\arrow[line width=1.5pt]{<}}}}] (0,0) to node [pos =0.7, above,sloped] {$\le(\begin{array}{cc} 0&1\\ -1&0\end{array}\ri)$} (180:2);
\end{tikzpicture}}
\end{center}
\caption{The jumps of the Airy Riemann-Hilbert problem.}
\label{jumpM}
\end{figure}
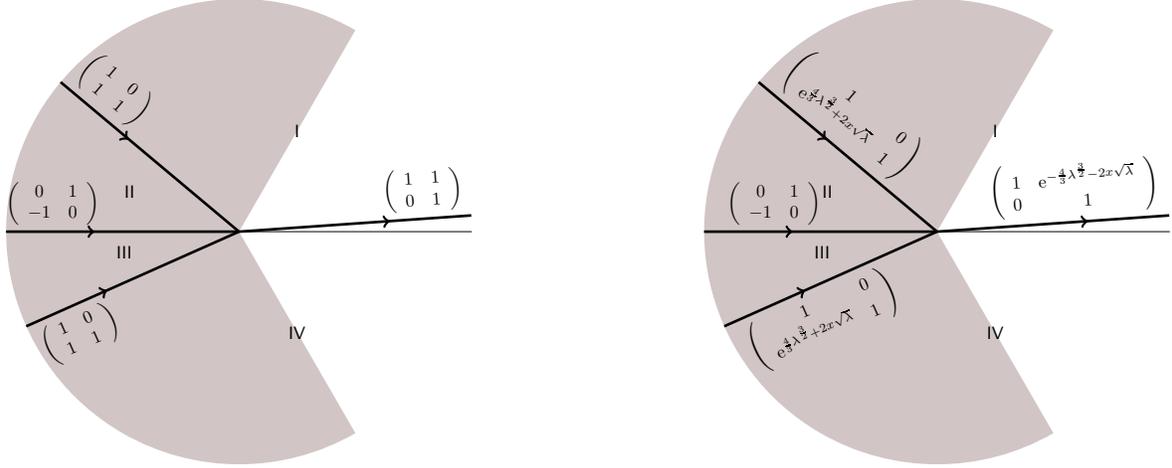

The RHP \ref{RHPgamma} for $\mathbf d_0 \equiv 1$ corresponds to an explicitly solvable problem involving Airy functions: we call it the {\it bare} solution. This is also the solution of the Painlev\'e hierarchy \ref{RHPP1N} with $N=1$  and $t_3 = \frac 23$.
\bd 
\label{Airyparametrix}
Let  $\omega:= {\rm e}^{2i\pi/3}$ and $\mathcal A(\z)$ be the matrix satisfying the jumps indicated in Fig. \ref{jumpM} and such that  
\begin{equation}\label{defA}
   \mathcal A(\zeta)=
   \sqrt{2\pi}e^{-\frac{\pi i}{12}} \times 
   \le\{
   \begin{array}{ll}
        \le[\begin{array}{cc} 
            \Ai(\zeta) & \Ai(\omega^2\zeta) \\
            \Ai'(\zeta) & \omega^2\Ai'(\omega^2\zeta)
        \end{array}\ri] {{\rm e}^{\frac {-i\pi }{6} \s_3}}, &  \mbox{for $\zeta\in I$,}
\\[10pt]
        \le[\begin{array}{cc} 
          -\omega  \Ai(\omega \zeta) & \Ai(\omega^2\zeta) \\
           -\omega^2  \Ai'(\omega \zeta) & \omega^2\Ai'(\omega^2\zeta)
        \end{array}\ri] {{\rm e}^{\frac {-i\pi }{6} \s_3}}, &  \mbox{for $\zeta\in II$,}
\\[10pt]
        \le[\begin{array}{cc} 
        -\omega^2    \Ai(\omega^2\zeta) &-\omega^2 \Ai(\omega\zeta) \\
           -\omega \Ai'(\omega^2\zeta) & -\Ai'(\omega\zeta)
        \end{array}\ri] {{\rm e}^{\frac {-i\pi }{6} \s_3}}, &  \mbox{for $\zeta\in III$,}
\\[10pt]
        \le[\begin{array}{cc} 
            \Ai(\zeta) &-\omega^2 \Ai(\omega\zeta) \\
            \Ai'(\zeta) & -\Ai'(\omega\zeta)
        \end{array}\ri] {{\rm e}^{\frac {-i\pi }{6} \s_3}}, &  \mbox{for $\zeta\in IV$,}
     \end{array}
     \ri.
\end{equation}
where the four regions are separated by the rays ${\rm e}^{i\theta_{0,\pm}} \R_+$ and $\R_-$ with 
the angles $\theta_{0,\pm}$  in the ranges
\be
\label{rangeAiry}
\theta_0 \in \le(-\frac \pi 3, \frac \pi 3\ri), \qquad \theta_{1} \in \le(\frac {\pi} 3, \pi \ri), \qquad
\theta_{-1} \in \le(-\pi, -\frac \pi 3\ri).
\ee
\ed
The matrix $M$ has the same asymptotic expansion  in each of the sectors I--IV (see, e.g. \cite{DKMVZ}); 
\bea\label{asymA}
        \mathcal A (\zeta) \! \sim \! 
&\&   {\rm e}^{-\frac {i\pi}4 \s_3}  \zeta^{-\frac{\sigma_3}{4}}\frac{\1+i\s_1}{\sqrt 2}\!\!
\left[
\1 +\sum_{k=1}^\infty\frac{1}{2}\left(\frac{2}{3}\zeta^{3/2}\right)^{\!\!-k} \!\!
            \le[\begin{array} {cc}
                (-1)^k(u_k+r_k) & i (u_k-r_k) \\
             -i (-1)^k(u_k-r_k) & u_k+r_k
            \end{array}\ri]
       \right] 
       e^{-\frac{2}{3}\zeta^{3/2}\sigma_3},
\eea
\begin{equation}\label{definition: sktk}
        u_k=\frac{\Gamma(3k+1/2)}{54^k k! \Gamma(k+1/2)},\qquad
        r_k=-\frac{6k+1}{6k-1}u_k,\qquad\mbox{for $k\geq 1$.}
\end{equation}
The matrix $\mathcal A$ solves the Airy equation (in matrix form) 
\be
\frac {\d}{\d \z}\mathcal A(\zeta) = \le[
\begin{array}{cc}
0 & 1\\
\zeta & 0
\end{array}
\ri]\mathcal A(\zeta).
\label{AiryODEmatrix}
\ee
Now we define 
\bea
\Gamma_0(\l ;x)  &\& := {\rm e}^{-\frac {i\pi}4 \s_3} \mathcal A(\l+ x) {\rm e}^{\le(\frac 23 \l^{\frac 32} +  x \sqrt{\l} \ri)\s_3},\label{Gamma0}\\ 
\Psi_0(\l;x)&\&  =:  {\rm e}^{-\frac {i\pi}4 \s_3} \mathcal A(\l+ x) = \G_0(\l;x)  {\rm e}^{-\le(\frac 23 \l^{\frac 32} +  x \sqrt{\l} \ri)\s_3}.
\eea
The matrix  $\Gamma_0$ provides the explicit solution of the Riemann--Hilbert problem \ref{RHPgamma} for $n=0$. 
Using the property \eqref{asymA}  one can verify directly that 
\bea
\Gamma_0(\l;x) = \l^{\frac {-\s_3}4} \frac { \1  + i\s_1}{\sqrt{2} } 
\le(\1  -\frac {x^2}4 \frac {\s_3}{\sqrt{\l}} - \frac {x}{4\l}{\s_2} + \mathcal O(\l^{-\frac 32 })\ri) 
\qquad  \qquad \hbox { as }  \l \to \infty.\label{G0infty}
\eea
By construction, $\mathcal A(\zeta)$ solves a Riemann--Hilbert problem with jumps indicated in the left pane of Fig. \ref{jumpM}. Consequently one can check that $\G_0(\l;x)$ solves the Riemann--Hilbert problem \ref{RHPgamma} for $n=0$. The only point worth remarking is that the jump contours  of $\mathcal A$ should be preemptively translated by $x$ so that the jump contours of $\G_0$ coincide exactly with rays issuing from the origin. 
\paragraph{The initial tau function.}
The isomonodromic tau function of the RHP for $n=0$ (which is  a function solely of $x$) is computed directly from the 
formula \eqref{JapTau} (using \eqref{G0infty}) which yields directly

\be
\label{initialtau}
\pa_x\ln \tau_0(x) = -\frac {x^2}{4} \ \ \Rightarrow \ \ \tau_0(x) = {\rm e}^{-\frac {x^3}{12}}.
\ee
It is worth remarking that $\tau_0(x)$ is nowhere vanishing: this is a signal that the RHP \eqref{RHPgamma} for $n=0$ is always solvable. 
\subsection{The dressing: discrete Schlesinger transformations}
The goal of this section is to determine the change of the following one form on the deformation space $\t$;
\be
\label{Omega}
\Omega(\pa; [M(\t)]) := \int_{\Sigma} \Tr \le(
\Gamma_{0-}^{-1}(\l;\t) \Gamma_{0-}'(\l;\t) \pa M(\l;\t) M^{-1}(\l;\t)
\ri)\frac {\d \l}{2i\pi}
\ee
when $M(\l;\t)$ is replaced by $M_n(\l;\t, \vec \l, \vec \mu ):= D_-^{-1}(\l) M(\l;\t) D_+(\l)$ and $D$ given in \eqref{defD}.
Now, in our setting the one form \eqref{Omega} is the total differential of the logarithm of the tau function (see also equation \eqref{Bertotau}).
\begin{shaded}
\bt
\label{taudressing}
The effect of the dressing of the jump matrices $M_n(\l;\t, \vec \l, \vec \mu ):= D^{-1}(\l) M(\l;\t) D(\l)$ on the one-form \eqref{Omega} is given by
\be
\label{varOmega}
\Omega(\pa; [M_n(\t)]) - \Omega(\pa; [M(\t)])  = \pa \ln \big({\Delta(\vec \l,\vec \mu) \det \mathbb G}
\big) \ee
where $\Delta(\vec \l, \vec \mu)$ and 
the $n\times n$ matrix $\mathbb G$   ($n = n_1+n_2$) are given by 
\bea
\label{defDelta}
\Delta(\vec \l,\vec \mu):= \frac {\prod_{j=1}^{n_2} \l_j^\frac 1 4\prod_{j=1}^{n_1} \mu_j^\frac 1 4 }{  \ds\prod_{j<k\leq n_2} (\sqrt{\l_j} - \sqrt{\l_k})\prod_{j<k\leq n_1} (\sqrt{\mu_j} - \sqrt{\mu_k})\prod_{j\leq n_2, k\leq n_1} (\sqrt{\l_j} + \sqrt{\mu_k}) }
\\
\label{Gentries}
\mathbb G_{k,\ell} = \le\{
\begin{array}{cc}
\ds  \res{\lambda=\infty}  \frac{  \l^{\lfloor \frac{\ell -1}2\rfloor}\mathbf e_2^{\mathrm T}\G_0^{-1}(\l_k)G_{\infty}(\l){\mathbf e}_{((\ell - 1) \, {\rm mod}\, 2)+1 }}{(\l-\l_k)}   & 1\leq k \leq n_2
\\
\ds  \res{\lambda=\infty}  \frac{  \l^{\lfloor \frac{\ell -1}2\rfloor}\mathbf e_1^{\mathrm T}\G_0^{-1}(\mu_{k-n_2})G_{\infty}(\l){\mathbf e}_{((\ell - 1) \, {\rm mod}\, 2)+1 }}{(\l-\mu_{k-n_2})}   & n_2+1\leq k \leq n_1+n_2
\end{array}
\ri.
\eea 
\be
\label{Ginfty}
G_\infty(\l) :=  \G_0(\l)D(\l) \frac {\1- i\s_1} {\sqrt{2}}  \l^{\frac {\s_3}4} \le\{
\begin{array}{cc}
\l^{-k} & n=2k\\
\le[
\begin{array}{cc}
\l^{-k-1} & 0 \\
0 & \l^{-k}
\end{array}
\ri] & n = 2k+1.
\end{array}
\ri.
\ee
Here $\pa$ denotes any derivatives with respect to $\t$ together with  the variables $\vec \l, \vec \mu$. 
\et
\end{shaded}

A proof by induction can be extracted from \cite{JMU2} but we will provide a different one which relies upon prior work in \cite{BertolaCafasso5} in  Appendix \ref{proofSchles}. We point out that the setting of Theorem \ref{taudressing} is precisely the one relevant to the Riemann--Hilbert problem \ref{RHPgamma}, where $M(\l;\t)$ is the jump matrix of the Airy Riemann--Hilbert problem for \eqref{Gamma0}.
\subsection{Proof of the main theorems}
We now return to the original setting and $\Gamma_0$ as given in \eqref{Gamma0}; in this case
we can further  simplify $\det \mathbb G$ and see that it provides the main ingredient for the Witten-Kontsevich tau  integral \eqref{ZnK} and  extensions \eqref{ZnKcont}.

\bp
\label{propdetG}
The following formula holds 
\be
\label{GnK}
\det \mathbb G \propto
{\rm e}^{\frac 23\le( \sum \l_j^\frac 32  -  \sum \mu_k^\frac 32\ri)+ x\le( \sum_j \l_j^\frac 1 2 - \sum\mu_k^\frac 12 \ri) } 
\det \le[
\begin{array}{l} 
\le[\AA_0^{(k-1)}(\l_j+x)\ri]_{1\leq k \leq n\atop 
\l_j\in I \cup IV}
\\
\hline 
\le[\AA_1^{(k-1)}(\l_j+x)\ri]_{1\leq k \leq n\atop 
\l_j\in I\!I}
\\
\hline 
\le[ \AA_1^{(k-1)}(\mu_j+x)\ri] _{{1\leq k\leq n
\atop
\mu_j\in I\!I\!I \cup I\!V
}}
\\
\hline 
\le[\AA_2^{(k-1)}(\l_j+x)\ri]_{1\leq k \leq n\atop 
\l_j\in I\!I\!I}
\\
\hline 
\le[ \AA_2^{(k-1)}(\mu_j+x)\ri] _{{1\leq k\leq n
\atop
\mu_j\in I \cup I\!I
}}
\end{array} \ri]
\ee
where 
$\AA_s (\l) = \Ai(\omega^s \l)$, $\omega = {\rm e}^{\frac{2i\pi}3}$, and $I, I\!I, I\!I\!I, I\!V$ denote the regions depicted in Fig. \ref{jumpM} and the proportionality is up to a constant independent of $\vec \l, \vec \mu$.
\ep
The  proof  is an elementary but somewhat lengthy manipulation using the form \eqref{Gentries} of the entries of $\mathbb G$, the explicit form of the matrix $\Gamma_0$ \eqref{Gamma0},  column operations and the differential equation satisfied by the Airy functions. We postpone it to the Appendix \ref{belinproof}.
%
%
%
\subsubsection{ Proof of Theorems \ref{main2bis}, \ref{main}, \ref{main2}}
\label{proofmain}
We denote $\l_k = y_k^2\ ,\ \ \forall y_k\in \{\Re y>0\}$ and $\mu_\ell = y_\ell^2 \ ,\ \ \ \forall y_\ell\in  \{\Re y<0\}$. 
Since the roots we use are all principal, we have
\be
{\rm e}^{\frac 23 \l_k^\frac 32 + x \l_k^\frac 1 2 } = 
{\rm e}^{\frac 23 y_k^ 3 + x y_k}\ ,\ \ \ \ 
{\rm e}^{-\frac 23 \mu_\ell^\frac 32 - x \mu_\ell^\frac 1 2 } = 
{\rm e}^{\frac 23 y_\ell^ 3+  x y_\ell}\ .
\ee
Then the determinant in  \eqref{GnK} becomes precisely the same as the determinant  in \eqref{ZnKcont} when written in terms of the $y_j$'s,  
 while $\Delta (\vec \l, \vec \mu)$  reduces to
\be
\Delta (\vec \l,\vec \mu) \mathop{=}^{\eqref{defDelta}} \pm   \frac {\prod_{j=1}^{n} \sqrt{y_j}}{\prod_{j<k} (y_j-y_k)}
\ee 
 up to an inessential sign.
We want to apply Theorem \ref{taudressing}; in this case
the jump matrix $M_n$ depends on $\vec \l, \vec \mu$ and $x$ only, and $\Omega(\pa;[M_n]) = \pa\ln \tau_n(x;\vec \l, \vec \mu )$, 
$\Omega(\pa_x ;[M]) = \pa_x \ln \tau_0(x) = -  \frac {x^2}4$ (see \eqref{initialtau}).
From Theorem \ref{taudressing} we obtain 
\be
\pa \ln \frac{\tau_n(x;\vec \l, \vec \mu )}{\tau_0(x)} = \pa \ln ((\det \mathbb G) \Delta(\vec \l;\vec \mu))  
 \mathop{=}^{\eqref{ZnKcont}} \pa \ln Z_n(x;\vec \l, \vec \mu).
\ee
The isomonodromic tau function $\tau_n$ is defined up to a multiplicative constant and therefore we can claim (using the expression   for $\tau_0(x)$ in \eqref{initialtau}) 
\be
\tau_n(x;\vec \l, \vec \mu ) = {\rm e}^{-\frac{x^3}{12}} Z_n(x;\vec \l, \vec \mu ).
\ee
The proof is now complete. \QED
 \br
\label{vars} 
The variables $\tau_n$ depend on, can be denoted as $\vec \l, \vec \mu$ or by $y_j = \sqrt \l_j, y_k = -\sqrt{\mu_k}$ (in the right/left half-planes of the $y$--plane). Furthermore, since the determinant in \eqref{GnK} is split into blocks depending on the index $\nu$ in $\Ai_\nu$, we can equivalently denote the dependence as  $\tau_n(x; \mathcal Y^{(0)}, \mathcal Y^{(1)}, \mathcal Y^{(2)})$. This is the way it was presented in the statement of Theorem \ref {main2bis}. 
 \er
\subsection{Approximation of  tau functions of the first  Painlev\'e hierarchy: proof of Theorem \ref{ThmP1N}}
\label{proofP1N}

We start with the following specializations of  Riemann--Hilbert problem \ref {RHPP1N}   as indicated below.

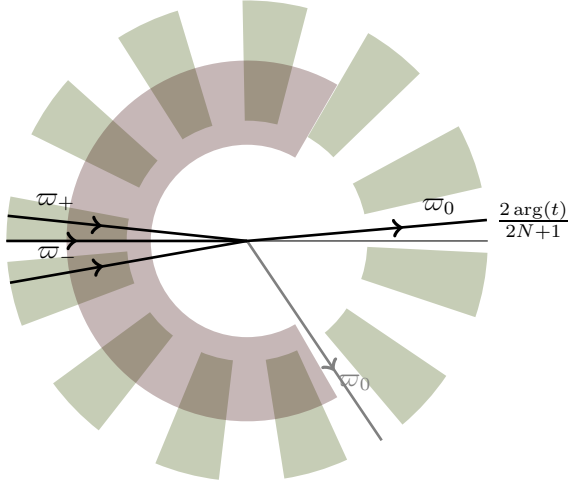
\begin{figure}
\begin{minipage}[c]{0.45\textwidth}
\begin{tikzpicture}[scale=1.6]
\foreach \y in {5}
{
\foreach \x in {-4.75,-5.75,-3.75,-2.75,-1.75,-0.75,0.25, 1.25, 2.25,3.25,4.25,5.25}
{
\foreach \angle in {360/(2*11+1)}
{
\fill [line width=0pt, fill= white!9!black!70!red!70!green, fill opacity = 0.3]  (0,0) to ({2*\x*\angle+\y}:2) 
arc ({max(2*\x*\angle + \y ,-180)}:{min({2*\x*\angle+\angle+\y} ,180)}:2);
};
};
\draw (0,0) to (\y:2) node [right] {$\frac {2\arg(t)}{2N+1}$};
};
\fill [white] (0,0) circle [radius =1];

\fill [line width=0pt, fill= white!7!black!80!red, fill opacity = 0.3]  (0,0) to (60:1.5) 
arc (60:300:1.5);
\fill [white] (0,0) circle [radius =0.8];
\draw (-2,0)--(2,0);

\draw [ line width = 1pt, gray, postaction={decorate,decoration={markings,mark=at position 0.65 with {\arrow[line width=1.5pt]{>}}}}] (0,0) to node [pos =0.8, above] {$\varpi_0$} (-56:2);
\draw [ line width = 1pt, postaction={decorate,decoration={markings,mark=at position 0.65 with {\arrow[line width=1.5pt]{>}}}}] (0,0) to node [pos =0.8, above] {$\varpi_0$} (5:2);
\draw [ line width = 1pt, postaction={decorate,decoration={markings,mark=at position 0.65 with {\arrow[line width=1.5pt]{<}}}}] (0,0) to node [pos =0.8, above] {$\varpi_+$} (174:2);
\draw [ line width = 1pt, postaction={decorate,decoration={markings,mark=at position 0.65 with {\arrow[line width=1.5pt]{<}}}}] (0,0) to node [pos =0.8, above] {$\varpi_-$} (190:2);
\draw [ line width = 1pt, postaction={decorate,decoration={markings,mark=at position 0.75 with {\arrow[line width=1.5pt]{<}}}}] (0,0) to  (180:2);

\end{tikzpicture}
\end{minipage}
\hfill
\begin{minipage}[c]{0.5\textwidth}
\caption{The jumps of the Riemann--Hilbert problem \ref{RHPP1Nspecial}. In the example $N = 11$ and the width of each darker sector is $2\pi/23$. 	
The rays $\varpi_\pm$ must extend to infinity within the sector shaded in both  hues, while $\varpi_0$ within the white sectors. In this example there are five  choices for  the ray $\varpi_{0}$ and four for each $\varpi_{\pm}$. Each choice determines a particular solution of the equation \eqref{P1N} of the  $P1_N$ hierarchy. 
In the example above (which is relevant to the setting of Theorem \ref{ThmP1N}),  shifting  $\arg t$ by $ \pm  \pi $ one of the two dark sectors adjacent to $\R_-$ disappears on the second sheet of $\sqrt{\l}$ and  one of the rays $\varpi_\pm$ is pinched. 
If we choose $\varpi_0$ as indicated by the lighter shade, then we can rotate $\arg t$ up to a smaller angle than $-\pi$ because the ray $\varpi_0$ will be forced to move in the sector $\mathcal S_1$, but we can still rotate up to $\pi$. In general, 
the reader can convince oneself that the  minimum amplitude of rotation of $\arg t$ is indeed $\pi$ in the positive and/or negative direction, and thus all these solutions of the hierarchy converge to the Airy parametrix exponentially fast  as $|t|\to 0$ and $\arg(t)$ in a sector of width at least $\pi$ that contains the positive real $t$--axis. 
}
\label{Sectors}
\end{minipage}
\end{figure}

\begin{problem}
Choose three integers $k_+, k_0 , k_-\,\,\in \le\{-\le\lfloor\frac  {N-1} 2\ri\rfloor,\dots,\le\lfloor\frac  {N-1} 2\ri\rfloor\ri\}$ with $k_+>k_-, \ k_+\geq k_0\geq k_-$ and 
specialize the Riemann--Hilbert problem \ref{RHPP1N} 
to the case $s_{2k_0} =1, s_{2k_\pm\pm1 } =- 1$. 
Furthermore set $t_1=x,\  t_3 = \frac 2 3, \  t_{2N+1}=t$ and all other $t_j=0$. Explicitly, the jump matrices read (with the rays oriented as in Fig. \ref{Sectors})
\label{RHPP1Nspecial}
\bea 
M (\l) = \le\{\begin{array}{lc}
\1 + {\rm e}^{-2\vartheta(\l;t,x)   }\s_+  &  \l \in  \varpi_0 := {\rm e}^{i\theta_0} \R_+
\\[8pt]
\1 +{\rm e}^{2\vartheta(\l;t,x) }\s_-    & \l \in  \varpi_\pm := {\rm e}^{i\theta_{\pm} } \R_+
 \\[8pt]
i\s_2  & \l\in \R_-
\end{array}\ri.
\label{Jump2}
\\
\vartheta(\l;t,x) = t \l^{\frac {2N+1}{2}}+ \frac 23 \l^\frac 32 +  x  \l^\frac 1 2
\label{varthetaspecial}
\eea
where the ray $\varpi_0 = {\rm e}^{i\theta_{0}}\R_+$  is such that  $\Re \l^\frac 32 >0< \Re t\l^{\frac {2N+1}2}$, and the two rays $\varpi_\pm = {\rm e}^{i\theta_{\pm }}\R_+$ are such that $\Re \l^\frac 32 <0> \Re t\l^{\frac {2N+1}2}$ (Fig. \ref{Sectors} for example). Namely we must have \eqref{rangeAiry} 
and 
\be
&\& \theta_0  \in \mathcal J_0(k_0, t):=\le(-\frac \pi {2N+1}, \frac \pi {2N+1} \ri) + \frac {4k_0\pi}{2N+1} -\frac {2 \arg (t)}{2N+1}\ ,\ \ k_0\in \Z
\nonumber \\ 
\label{rangeP1N}
&\& \theta_{\pm}   \in \mathcal J_\pm(k_\pm, t):= \le(-\frac \pi {2N+1}, \frac \pi {2N+1} \ri) + \frac {(4k_\pm \pm 2)\pi}{2N+1} -\frac {2 \arg (t)}{2N+1}\ ,\ \ k_\pm \in \Z
\ee

\end{problem}
\bp \label{propP1toAiry}\mbox{}\\
\noindent {\bf [1]} Let $\G(\l; t, x)$ be the solution of the Riemann--Hilbert problem \eqref{RHPP1Nspecial} with a choice of integers $\vec k=(k_0,k_+,k_-)$ such that 
\be
\mathcal J_0(k_0,1) \cap \le(\frac {-\pi} 3, \frac \pi 3\ri)\neq \emptyset
 \qquad \ \mathcal J_+(k_+,1) \cap 
\le( \frac \pi 3, \pi \ri)\neq\emptyset\ \qquad
\mathcal J_-(k_-,1) \cap\le( -\pi, -\frac \pi 3 \ri)\neq \emptyset.
\ee
 Then for $x$ ranging in a compact set the solution $\G$ is analytic for $t$ in a sector $\{|t|<r,\ \ \arg(t)\in (a,b)\}$ containing $\arg t=0$, that depends on the choice of $\vec k$ and has width at least $\pi$. For the case $k_0=0$ the sector contains the sector $\arg (t) \in (-\pi, \pi)$. 
 
\noindent {\bf [2]}  This solution, for $t\rightarrow 0$, converges to the  Airy parametrix \eqref{Gamma0} and also  its tau function $\tau_{P1_N}(\t)$ defined by \eqref{defTauP1N} (with $t_1=x, t_{2N+1}=t$ and all other $t_j=0$),  converges to ${\rm e}^{\frac{x^3}{12}.}$

\noindent {\bf [3]} The derivatives of arbitrary order with respect to $t_{2N+1} =t, t_1 =x$ of $\tau_{P1_N}(x,t)$ also converge as $|t|\to 0$ in the same sector to the derivatives of the topological solution. 
\ep
\noindent {\bf Proof.} 
{\bf [1]} The matrix  $\G_0 $ in Def. \ref{Airyparametrix} solves the RHP  \eqref{RHPgamma} with $\mathbf d_0 \equiv 1$. The  three rays $\omega \R_+$  can be rotated to rays $\varpi_j = {\rm e}^{i\theta_j} \R_+$,  $j= 0,\pm 1$ within the range \eqref{rangeAiry}. 

Indeed, within these ranges  the  jump matrices in Fig. \eqref{jumpM} are of the form $\1 + \mathcal O(|\l|^{-\infty})$ as $|\l|\to\infty$  since the function  ${\rm e}^{-\frac 23 \l^\frac 32}$ is decaying  along $\varpi_0$ and ${\rm e}^{\frac 23 \l^\frac 32}$ is decaying along $\varpi_\pm$.

On the other hand the well-posedness of the Riemann--Hilbert problem \ref{RHPgamma} for $\G$ requires that the rays satisfy \eqref{rangeP1N} 
so that ${\rm e}^{-t \l^{\frac {2N+1}2}}$ is decaying along $\varpi_0$ and  ${\rm e}^{t \l^{\frac {2N+1}2}}$ is decaying along $\varpi_\pm$, see Fig. \ref{Sectors}.

\begin{figure}

\begin{tikzpicture}[scale=1.5]
\foreach \x in { -2.75,-1.75,-0.75,0.25, 1.25, 2.25}
{
\foreach \y in {0}
{
\fill [line width=0pt, fill= white!9!black!70!red!70!green, fill opacity = 0.3]  (0,0) to ({2*\x*360/13+\y}:2) 
arc ({max(2*\x*360/13 + \y ,-180)}:{min({2*\x*360/13+360/13+\y} ,180)}:2);
};
};
\fill [white] (0,0) circle [radius =1];
\fill [line width=0pt, fill= white!7!black!80!red, fill opacity = 0.3]  (0,0) to (60:1.5) 
arc (60:300:1.5);
\fill [white] (0,0) circle [radius =0.8];
\draw (-2,0)--(2,0);

\draw [ line width = 1pt, postaction={decorate,decoration={markings,mark=at position 0.65 with {\arrow[line width=1.5pt]{>}}}}] (0,0) to node [pos =0.7, above] {$\varpi_0$} (0:2);
\draw [ line width = 1pt, postaction={decorate,decoration={markings,mark=at position 0.65 with {\arrow[line width=1.5pt]{<}}}}] (0,0) to node [pos =0.6, above] {$\varpi_+$} (180-360/13*1.5:2);
\draw [ line width = 1pt, postaction={decorate,decoration={markings,mark=at position 0.65 with {\arrow[line width=1.5pt]{<}}}}] (0,0) to node [pos =0.66, above] {$\varpi_-$} (180+360/13*1.5:2);
\draw [ line width = 1pt, postaction={decorate,decoration={markings,mark=at position 0.65 with {\arrow[line width=1.5pt]{<}}}}] (0,0) to  (180:2);

\foreach \r in {-3, -2,-1,0,1,2,3}
{
\begin{scope}[rotate= \r*360/13*2]
\foreach \x in 
{  ( 1.6271215189, -.3290135864 ) ,  ( 1.6271215189, .3290135864 ) ,  ( 1.6150741589, -.2686943171 ) ,  ( 1.6150741589, .2686943171 ) ,  ( 1.6069614059, -.2173990609 ) ,  ( 1.6069614059, .2173990609 ) ,  ( 1.6011495977, -.1704748096 ) ,  ( 1.6011495977, .1704748096 ) ,  ( 1.5970044747, -.1261062714 ) ,  ( 1.5970044747, .1261062714 ) ,  ( 1.5942081924, -0.833097940e-1 ) ,  ( 1.5942081924, 0.833097940e-1 ) ,  ( 1.5925884348, -0.414389072e-1 ) ,  ( 1.5925884348, 0.414389072e-1 ) ,  ( 1.5920575290, 0. ) 
}
{
\draw [fill]\x circle[radius=0.5pt];
};
\end{scope}};
\foreach \r in {-2,-1,0,1,2,-3}
{
\begin{scope}[rotate= 360/13+ \r*360/13*2]
\foreach \x in 
{ ( 1.6271215189, -.3290135864 ) ,  ( 1.6271215189, .3290135864 ) ,  ( 1.6150741589, -.2686943171 ) ,  ( 1.6150741589, .2686943171 ) ,  ( 1.6069614059, -.2173990609 ) ,  ( 1.6069614059, .2173990609 ) ,  ( 1.6011495977, -.1704748096 ) ,  ( 1.6011495977, .1704748096 ) ,  ( 1.5970044747, -.1261062714 ) ,  ( 1.5970044747, .1261062714 ) ,  ( 1.5942081924, -0.833097940e-1 ) ,  ( 1.5942081924, 0.833097940e-1 ) ,  ( 1.5925884348, -0.414389072e-1 ) ,  ( 1.5925884348, 0.414389072e-1 ) ,  ( 1.5920575290, 0. ) }
{
\draw [fill,red!80!black]\x circle[radius=0.5pt];
};
\end{scope}};
\node at (60:0.5) {I};
\node at (160:0.5) {II};
\node at (-160:0.5) {III};
\node at (-60:0.5) {IV};
\end{tikzpicture}
\hfill 
\begin{tikzpicture}[scale=1.5]
\foreach \x in { -2.75,-1.75,-0.75,0.25, 1.25, 2.25}
{
\foreach \y in {0}
{
\fill [line width=0pt, fill= white!9!black!70!red!70!green, fill opacity = 0.3]  (0,0) to ({2*\x*32.72+\y}:2) 
arc ({max(2*\x*32.72 + \y ,-180)}:{min({2*\x*32.72+32.72+\y} ,180)}:2);
};
};
\fill [white] (0,0) circle [radius =1];
\fill [line width=0pt, fill= white!7!black!80!red, fill opacity = 0.3]  (0,0) to (60:1.5) 
arc (60:300:1.5);
\fill [white] (0,0) circle [radius =0.8];
\draw (-2,0)--(2,0);

\draw [ line width = 1pt, postaction={decorate,decoration={markings,mark=at position 0.65 with {\arrow[line width=1.5pt]{>}}}}] (0,0) to node [pos =0.7, above] {$\varpi_0$} (0:2);
\draw [ line width = 1pt, postaction={decorate,decoration={markings,mark=at position 0.65 with {\arrow[line width=1.5pt]{<}}}}] (0,0) to node [pos =0.7, above] {$\varpi_+$} (180-32.72/2:2);
\draw [ line width = 1pt, postaction={decorate,decoration={markings,mark=at position 0.65 with {\arrow[line width=1.5pt]{<}}}}] (0,0) to node [pos =0.7, above] {$\varpi_-$} (180+32.72/2:2);
\draw [ line width = 1pt, postaction={decorate,decoration={markings,mark=at position 0.65 with {\arrow[line width=1.5pt]{<}}}}] (0,0) to  (180:2);

\foreach \r in {-2,-1,0,1,2}
{
\begin{scope}[rotate= \r*32.72*2]
\foreach \x in 
{ ( 1.7699360275, -.4252457742 ),  ( 1.7699360275, .4252457742 ),  ( 1.7569389915, -.3467034475 ),  ( 1.7569389915, .3467034475 ),  ( 1.7482683374, -.2801947396 ),  ( 1.7482683374, .2801947396 ),  ( 1.7420990216, -.2195341113 ),  ( 1.7420990216, .2195341113 ),  ( 1.7377210881, -.1623002286 ),  ( 1.7377210881, .1623002286 ),  ( 1.7347784130, -.1071772184 ),  ( 1.7347784130, .1071772184 ),  ( 1.7330778446, -0.532981766e-1 ),  ( 1.7330778446, 0.532981766e-1 ),  ( 1.7325210936, 0. )}
{
\draw [fill]\x circle[radius=0.5pt];
};
\end{scope}};
\foreach \r in {-2,-1,0,1,2,-3}
{
\begin{scope}[rotate= 32.72+ \r*32.72*2]
\foreach \x in 
{ ( 1.7699360275, -.4252457742 ),  ( 1.7699360275, .4252457742 ),  ( 1.7569389915, -.3467034475 ),  ( 1.7569389915, .3467034475 ),  ( 1.7482683374, -.2801947396 ),  ( 1.7482683374, .2801947396 ),  ( 1.7420990216, -.2195341113 ),  ( 1.7420990216, .2195341113 ),  ( 1.7377210881, -.1623002286 ),  ( 1.7377210881, .1623002286 ),  ( 1.7347784130, -.1071772184 ),  ( 1.7347784130, .1071772184 ),  ( 1.7330778446, -0.532981766e-1 ),  ( 1.7330778446, 0.532981766e-1 ),  ( 1.7325210936, 0. )}
{
\draw [fill,red!80!black]\x circle[radius=0.5pt];
};
\end{scope}};

\node at (60:0.5) {I};
\node at (172:0.8) {II};
\node at (-172:0.8) {III};
\node at (-60:0.5) {IV};
\end{tikzpicture}
\caption{The points $\mu_k$ (red) and $\l_j$ (black). $N=6$ (left) and $N=5$ (right).}
\label{figP1Ntronquee}
\end{figure}
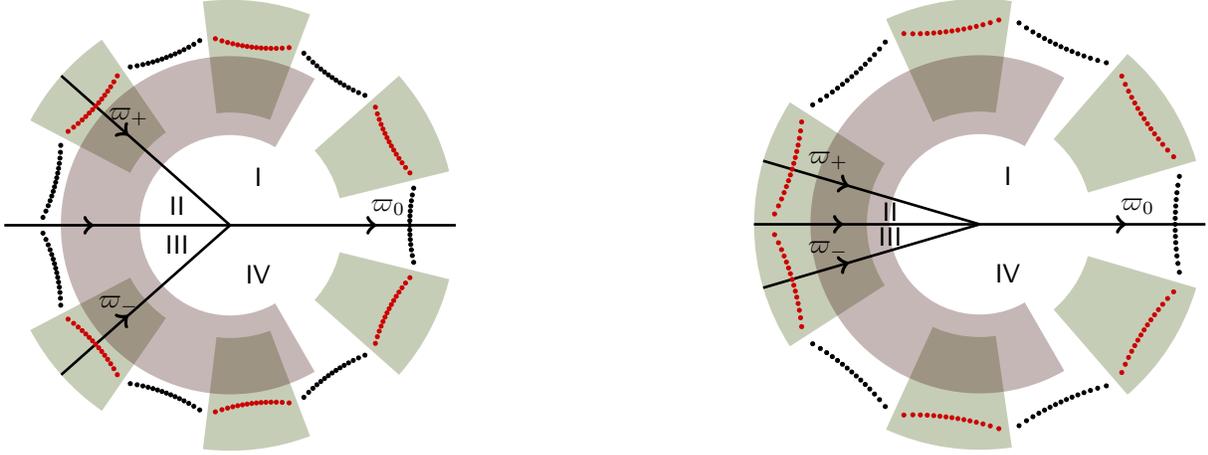
In general there are several possible choices of $k_0, k_\pm $ in \eqref{rangeP1N} that satisfy both ranges \eqref{rangeAiry}, \eqref{rangeP1N}; for a given choice of rays, the conditions will remain satisfied within a certain maximal open sector in the $t$--plane;
it should be noted that different choices of $k_{0,\pm}$ lead to {\it inequivalent} Riemann--Hilbert problems and solutions of the first Painlev\'e\ hierarchy (see below); in particular the poles of the corresponding Painlev\'e\ transcendent depend on this choice. 

  Let us assume now that we make one such choice of sector and consider the RHP for the matrix $\mathcal E$ with jumps on $\varpi_{0,\pm 1}$ as follows
\be
\mathcal E_+(\l; t,x) = \mathcal E_-(\l; t,x)  \le(\1 + 
{\rm e}^{-\frac  23 \l^\frac 32 - x \sqrt\l} \le({\rm e}^{-t \l ^{\frac {2N+1}2}} -1\ri) \G_0(\l;x) \s_+\G_0^{-1}(\l;x)\ri)\ ,\ \ \l \in \varpi_0
\label{JE1}
\\
\mathcal E_+(\l; t,x) = \mathcal E_-(\l; t,x)  \le(\1 + 
{\rm e}^{\frac  23 \l^\frac 32 + x \sqrt\l} \le({\rm e}^{t \l ^{\frac {2N+1}2}} -1\ri) \G_0(\l;x) \s_-\G_0^{-1}(\l;x)
\ri)\ ,\ \ \l \in \varpi_\pm
\label{JE2}\\
\mathcal E(\l;t,x) = \1 + \mathcal O(\l^{-1}) \ ,\ \ \l\to \infty.
\ee
Along the rays in the chosen sector, the matrices $\G_0, \G_0^{-1}$ (Airy parametrix) in \eqref{JE1}, \eqref{JE2} remain bounded, uniformly with respect to $x$ ranging in a compact set.
Furthermore, within the chosen  sector, we can send $|t|\to 0$ and the jump matrices will converge to the identity matrix in all $L^p$ norms, $1\leq p \leq \infty$, uniformly with respect to  $x$ in compact sets:
for example on $\varpi_+$ the function $ {\rm e}^{\frac  23 \l^\frac 32 + x \sqrt\l} \le({\rm e}^{t \l ^{\frac {2N+1}2}} -1\ri)$ belongs to all $L^p(\varpi_+, |\d\l|)$ because  ${\rm e}^{\frac  23 \l^\frac 32 + x \sqrt\l}$ does, and $\le({\rm e}^{t \l ^{\frac {2N+1}2}} -1\ri)$ is bounded.
 Consequently the matrix $\mathcal E$ converges to the identity as $|t|\to 0$ in the given sector and  has an expansion near $\l=\infty$ of the form
\be
\label{328}
\mathcal E(\l; t,x) = \1 + \frac 1 \l \mathcal E_1(t,x) + \mathcal O(\l^{-2}).
\ee
Most importantly, for $|t|$ sufficiently small, the existence of the solution $\mathcal E$ (and its analyticity with respect to the parameters $t,x$) is guaranteed by standard arguments.
By construction of the jump relations \eqref{JE1}, \eqref{JE2}, the matrix $\mathcal E(\l; t,x) \G_0(\l;x)$ solves a Riemann--Hilbert problem with the same jumps as $\G(\l; t, x)$  but in a different gauge (see Remark \ref{remgauge}). By a {\it left} multiplication with  $\l$--independent matrix and by the uniqueness of the solution of the Riemann--Hilbert problem \ref{RHPgamma}, we deduce that 
\be
 \G(\l; t, x) =\le(\1 - (\mathcal E_1(t,x))_{12} \s_-\ri) \mathcal E(\l; t,x) \G_0(\l;x)\ .
\ee
The left multiplier is crafted so as to guarantee the same gauge as $\G$ at infinity (see Rem. \ref{remgauge}).
We conclude that $\G(\l; t,x)$ is analytic in the specified domain. 
The width of the sectors is explained by way of example in the caption of Fig. \ref{Sectors}.
\\
{\bf [2]} Since $\mathcal E\to \1$, we deduce that the tau function  for $\G(\l;t,x)$  defined by \eqref{defTauP1N}  converges to that of $\G_0$  as given in \eqref{initialtau}.\\ 
{\bf [3]} By the same argument, using \eqref{higherderivatives} we conclude that all derivatives of $\ln \tau_{P1_N}$ converge as $|t|\to 0$ within the common sector, to the same expression \eqref{higherderivatives} evaluated using the Airy parametrix $\Gamma_0$. These are \cite{BertolaDiDubrovin} precisely the derivatives of the topological solution of KdV (note that we are using a different normalization of the time $t$ from loc. cit.  but this is inconsequential to our discussion).\QED

\subsubsection{Equivalence to all orders of different solutions: proof of Thm. \ref{ThmP1N}$_{[3]}$. }
\label{sectP1stokes}
\begin{figure}
\begin{tikzpicture}[scale=1.6]
\foreach \x in {-3.75,-2.75,-1.75,-0.75,0.25, 1.25, 2.25,3.25}
{
\fill [line width=0pt, fill= white!9!black!70!red!70!green, fill opacity = 0.3]  (0,0) to (2*\x*24+9:2) 
arc ({max(2*\x*24 + 9,-180)}:{min(2*\x*24+24+9 ,180)}:2);
};
\fill [white] (0,0) circle [radius =1.1];

\fill [line width=0pt, fill= white!7!black!80!red, fill opacity = 0.3]  (0,0) to (60:1.5) 
arc (60:300:1.5);
\fill [white] (0,0) circle [radius =0.8];

\draw (-2,0)--(2,0);

\draw [ line width = 1pt, postaction={decorate,decoration={markings,mark=at position 0.65 with {\arrow[line width=1.5pt]{>}}}}] (0,0) to node [pos =0.8, above] {$\varpi_0$} (12:2);
\draw [ line width = 1pt, postaction={decorate,decoration={markings,mark=at position 0.65 with {\arrow[line width=1.5pt]{<}}}}] (0,0) to node [pos =0.8, above] {$\varpi_+$} (170:2);
\draw [ line width = 1pt, postaction={decorate,decoration={markings,mark=at position 0.65 with {\arrow[line width=1.5pt]{<}}}}] (0,0) to node [pos =0.8, above] {$\varpi_-$} (194:2);

\draw [ line width = 1pt, black!40!blue, postaction={decorate,decoration={markings,mark=at position 0.65 with {\arrow[line width=1.5pt]{<}}}}] (0,0) to node [pos =0.8, above] {$\wt \varpi_+$} (125:2);
\draw [ line width = 1pt, black!40!blue, postaction={decorate,decoration={markings,mark=at position 0.65 with {\arrow[line width=1.5pt]{<}}}}] (0,0) to node [pos =0.8, above] {$\wt \varpi_-$} (248:2);
\draw [ line width = 1pt, black!40!blue, postaction={decorate,decoration={markings,mark=at position 0.65 with {\arrow[line width=1.5pt]{>}}}}] (0,0) to node [pos =0.8, above] {$\wt \varpi_0$} (-43:2);

\end{tikzpicture}\hfill
\begin{tikzpicture}[scale=1.6]
\foreach \x in {-3.75,-2.75,-1.75,-0.75,0.25, 1.25, 2.25,3.25}
{
\fill [line width=0pt, fill= white!9!black!70!red!70!green, fill opacity = 0.3]  (0,0) to (2*\x*24+9:2) 
arc ({max(2*\x*24 + 9,-180)}:{min(2*\x*24+24+9 ,180)}:2);
};
\fill [white] (0,0) circle [radius =1.1];

\fill [line width=0pt, fill= white!7!black!80!red, fill opacity = 0.3]  (0,0) to (60:1.5) 
arc (60:300:1.5);
\fill [white] (0,0) circle [radius =0.8];

\draw (-2,0)--(2,0);
\begin{scope}[rotate=6]
\draw[line width = 1pt, postaction={decorate,decoration={markings,mark=at position 0.5 with {\arrow[line width=1.5pt]{>}}}}] (194:2) to (194:1)   arc (194:221:1);
 \draw[line width = 1pt, black!40!blue, postaction={decorate,decoration={markings,mark=at position 0.65 with {\arrow[line width=1.5pt]{>}}}}] (221:1) arc(221:242:1)
  to (242:2);
 \draw [dashed, xshift=-0.2pt,yshift=0.3pt, postaction={decorate,decoration={markings,mark=at position 0.35 with {\arrow[line width=1pt]{<}}}}] (0,0)--(221:1);
 \draw [ dashed,black!40!blue , postaction={decorate,decoration={markings,mark=at position 0.65 with {\arrow[line width=1pt]{>}}}}](0,0)--(221:1);
 \end{scope}
\begin{scope}[xscale=-1, rotate =-26]
\draw[line width = 1pt, postaction={decorate,decoration={markings,mark=at position 0.5 with {\arrow[line width=1.5pt]{<}}}}] (196:2) to (196:1) arc(196:221:1);
 \draw[line width = 1pt, black!40!blue, postaction={decorate,decoration={markings,mark=at position 0.6 with {\arrow[line width=1.5pt]{<}}}}] (221:1) arc(221:242:1)
 to (242:2);
 \draw [dashed, xshift=-0.2pt,yshift=0.3pt, postaction={decorate,decoration={markings,mark=at position 0.35 with {\arrow[line width=1pt]{>}}}}] (0,0)--(221:1);
 \draw [ dashed,black!40!blue , postaction={decorate,decoration={markings,mark=at position 0.65 with {\arrow[line width=1pt]{<}}}}](0,0)--(221:1);
\end{scope}
\begin{scope}[xscale=-1, rotate =170]
\draw[line width = 1pt, postaction={decorate,decoration={markings,mark=at position 0.5 with {\arrow[line width=1.5pt]{>}}}}] (196:2) to  node[pos=0.5, above] {$ \gamma_1$} (196:1)  arc(196:221:1) node [left] {$\Pi$} ;
 \draw[line width = 1pt, black!40!blue, postaction={decorate,decoration={markings,mark=at position 0.65 with {\arrow[line width=1.5pt]{>}}}}] (221:1) arc(221:242:1) to  node[pos=0.5, above] {$ \gamma_2$} (242:2);
 \draw [dashed, xshift=-0.2pt,yshift=0.3pt, postaction={decorate,decoration={markings,mark=at position 0.35 with {\arrow[line width=1pt]{<}}}}] (0,0)--(221:1);
 \draw [ dashed,black!40!blue , postaction={decorate,decoration={markings,mark=at position 0.65 with {\arrow[line width=1pt]{>}}}}](0,0)--(221:1);
\end{scope}
\end{tikzpicture}

\caption{Left: the jump contours $\varpi_{0,\pm}$ and $\wt \varpi_{0,\pm}$ (marked in different colors) of two  solutions $\Gamma, \wt \Gamma$ of the RHP \ref{RHPP1Nspecial} with different choices of the integers $(k_0,k_+, k_-)$ (specifically, $N=7$ and  $(k_+ = 3, k_0 = 0, k_- = -3)$ while $(\wt k_+ = 2, \wt k_0 = -1, \wt k_- = -2)$) .  Right: the jumps of their ratio $\mathcal E= \Gamma \wt \Gamma^{-1}$  after the contour deformation. On the dashed arcs the jumps cancel each other and hence $\mathcal E$ is continuous across them.}
\label{P1stokes}
\end{figure}
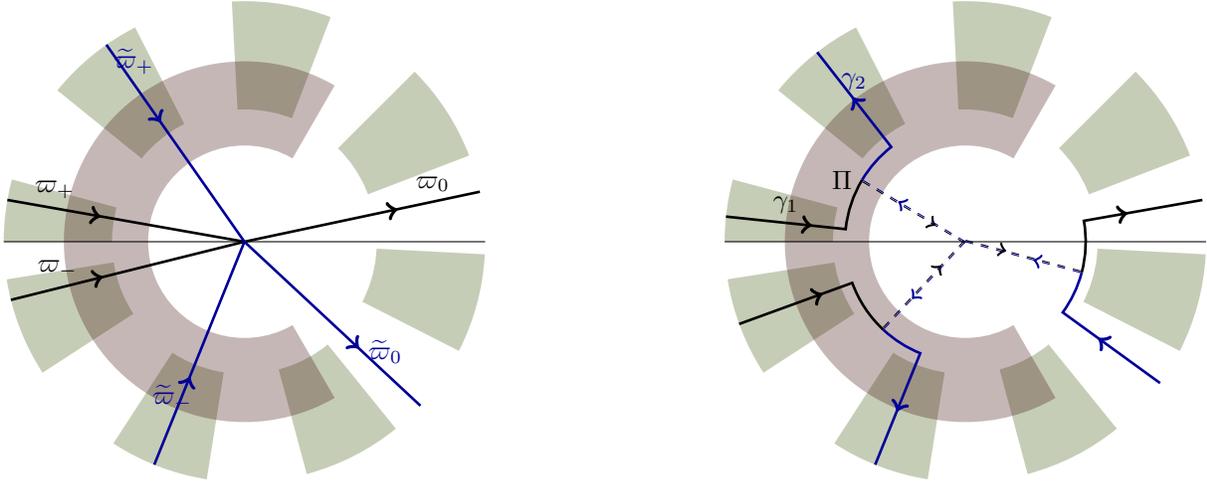

Proposition \ref{propP1toAiry} has already established Theorem \ref{ThmP1N}$_{[2]}$ and part of Theorem \ref{ThmP1N}$_{[3]}$. It remains to show that two solutions of the Riemann--Hilbert problem \ref{RHPP1Nspecial} (and the corresponding tau functions) with different choices of $\vec k=(k_0,k_+,k_-)$ differ by exponentially small terms as $|t|\to 0$ as long as the corresponding sectors appearing in the Proposition \ref{propP1toAiry}  have non-empty overlap. See Fig. \ref{P1stokes} illustrating  a typical such setup.

The proof is a simple application of perturbation analysis of Riemann--Hilbert problems; the ratio of two solutions with different choices $\vec k_1, \vec k_2$ has a jump which approaches the identity in any $L^p$ ($1\leq p \leq \infty$) at an exponential rate in $\frac 1{|t|^\sharp}$, with a power law that we are going to compute. 

We need to analyze the signs of the real part of the phase function $\vartheta$ \eqref{varthetaspecial} as $|t|\to 0$. 
Treating the term with $t$ in $\vartheta$ as a perturbation, it is clear that for $|t|$ sufficiently small the signs of $\Re \vartheta(\l;t,x)$ are dominated by those of $f_0 = \Re (\frac 23 \l^\frac 23  + x\sqrt \l)$ in any fixed compact set in the $\l$--plane. 
 Let us fix  a bounded domain for $x$: $|x|< K$.  
 
We are free to deform the jump contours of the RHP \ref{RHPP1Nspecial} as we wish as long as the asymptotic directions at infinity satisfy the appropriate conditions, we shall deform them in a way that we explain below.

\paragraph{Contour deformation.} 
We consider $t\in \R_+$ for simplicity  because the steps are valid in a small sector.  For definiteness we only treat  $\varpi_+$, with similar considerations in the remaining cases. 
Let $\theta_1 := \frac {(2k_++1)2\pi}{2N+1}, \ {\theta_2} := \frac {(2\wt k_++1)2\pi}{2N+1}$ be the bisecants of the sectors visited by the jumps of the two solutions (refer to Fig. \ref{P1stokes}) and recall that by assumption we must have $\theta_{1,2} \in (\frac \pi 3, \pi)$.  
Along the two rays $\arg \l = \theta_{1,2}$ we have (here $r  = |\l|$)
\be
\Re \vartheta = - |t|\, r^{\frac {2N+1}2}  + \frac 2 3 r^\frac 32 \cos\le( \frac 3 2\theta_j\ri) + \Re x \sqrt\l < \sqrt r \le[- |t| r^N   -Q \frac 23 r + K \ri]
\label{est1}
\ee
where $Q = \min |\cos(\frac 32 \theta_j)|$ is a positive number because of the condition $\theta_j \in (\frac \pi 3, \pi)$.

Consider the rays $\gamma_j$ given by  $\arg \l = \theta_j$ $r\geq r_0 =  (Q/2)^\frac 1{N-1} |t|^{\frac {-1}{N-1}} $; along these rays the function ${\rm e}^{\vartheta}$ belongs to any $L^p$ ($1\leq p\leq \infty$) and the corresponding norm decays exponentially as $|t|\to 0$; in fact  from \eqref{est1}
\be
\sup_{\l\in \gamma} \Re \vartheta(\l) \leq
t^\frac {-3}{2N-2}\frac {Q^{\frac {N+1/2}{N-1}}}{2^\frac 1{N-1}}\le( - \frac 1 2 - \frac 23 + \frac {K |t|^\frac {1}{N-1}}{Q^\frac N{N-1} 2^\frac 3{2N-2}}\ri)=
t^\frac {-3}{2N-2}\frac {Q^{\frac {N+1/2}{N-1}}}{2^\frac 1{N-1}}\le( - \frac 76 + \frac {K |t|^\frac {1}{N-1}}{Q^\frac N{N-1} 2^\frac 3{2N-2}}\ri)
\ee
which clearly tends to $-\infty$ as $|t|\to 0$. Consider now the arc of circle $\Pi$, joining the two points $r_0{\rm e}^{i\theta_1}$ to $r_0{\rm e}^{i\theta_2}$; 
along this arc we have (with $\varphi = \arg \l$);
\be
\Re \vartheta  = - |t|\, r^{\frac {2N+1}2} \cos \le(\frac {2N+1}2\varphi \ri)  + \frac 2 3 r^\frac 32 \cos\le( \frac 3 2\varphi\ri) +\Re ( x \sqrt\l)
\leq t^\frac {-3}{2N-2}\frac {Q^{\frac {N+1/2}{N-1}}}{2^\frac 1{N-1}}\le(  - \frac 16 + \frac {K |t|^\frac {1}{N-1}}{Q^\frac N{N-1} 2^\frac 3{2N-2}}\ri)\cr
\label{est2}
\ee
where we have used that $\cos(\frac 32 \varphi)\leq -Q$ along the arc. Once more the $L^p$ norm of ${\rm e}^{\vartheta}$ along this arc is easily estimated to tend to zero exponentially. 
\paragraph{Exponential rate of convergence.}
Now, referring to Fig. \ref{P1stokes} we deform the rays $\varpi_+, \wt{\varpi_+}$ as indicated and consider the RHP for the matrix $\mathcal E$ with  jumps only on the union of the rays $\gamma_{1,2}$ and the arc $\Pi$ as shown in the figure and with the jump matrix given by  $\1 +  {\rm e}^{\vartheta(\l;t,x)}\wt \Gamma (\l;t,x) \s_-\wt \Gamma^{-1}(\l;t,x)$. We know from Prop. \ref{propP1toAiry} that $\wt \Gamma(\l;t,x)$ tends to the Airy parametrix as $|t|\to 0$ in a small sector around $\R_+$ uniformly with respect to $\l$ on the Riemann sphere and hence it remains bounded as $|t|\to 0$. 
The $L^p$ norms of ${\rm e}^{\vartheta}$ on the two rays are $\mathcal O\le({ \rm e}^{- \wt C |t|^{-\frac {3}{2N-2}}}\ri)$ while the $L^p$ norms on the arc $\Pi$ (whose length grows like $|t|^{-\frac {1}{N-1}}$), are all bounded by $\mathcal O\le ( 
|t|^{-\frac {1}{N-1}} { \rm e}^{- C |t|^{-\frac {3}{2N-2}}}\ri)$, where $C, \wt C$ are  positive constants that follow from the estimates \eqref{est1}, \eqref{est2}. 
In total the $L^p$ norm of ${\rm e}^{\vartheta}$ along the whole contour are bounded by 
\be
\le\|{\rm e}^{\vartheta} \ri\|_{L^p( \gamma_1 \cup \gamma_2 \cup \Pi, |\d\l|) } = \mathcal O\le(
t^{\frac {-1}{N-1}}  { \rm e}^{- C |t|^{-\frac {3}{2N-2}}}
\ri) ,\label{333}
\ee
uniformly with respect to $1\leq p\leq \infty$. By standard arguments \cite{Deift} on small norm Riemann Hilbert problems, we conclude that $\mathcal E$ converges to the identity on the Riemann sphere, at the same exponential rate \eqref{333}
Then, by the same argument used in Proposition \ref{propP1toAiry}, we conclude that 
$$
 \G(\l; t, x) =\le(\1 - (\mathcal E_1(t,x))_{12} \s_-\ri) \mathcal E(\l; t,x) \wt \G(\l;x)\
$$
(with $\mathcal E_1(t,x)$ similar as in \eqref{328}) 
 and hence $\wt \Gamma (\l;t,x)$, $\Gamma(\l, t,x)$ differ from each other by exponentially small terms as $|t|\to 0$ in a small sector around $\R_+$. In particular, the ratio of the corresponding tau functions defined via \eqref{defTauP1N} will also tend to unity exponentially fast as $|t|\to 0$; therefore, the asymptotic expansion of the logarithms of the  two tau functions in powers of $t$ will be identical in the overlapping sector. Since the estimates are uniform with respect to $x$ in a compact set (we used $|x|<K$ in the estimates \eqref{est1}, \eqref{est2}), the coefficients of these expansions must also be analytic in $x$ at least in the same domain.

\subsubsection{Pad\'e approximation: proof of Thm. \ref{ThmP1N}$_{[1]}$}
\label{proofP1N1}
The exponential function admits a Pad\'e\ approximation  of the form
\be
{\rm e}^{-z} = \prod_{j=1}^r \frac {a_j  -  z}{a_j+z}   + \mathcal O(z^{2r+1})  = \frac{P_r(z)}{P_r(-z)}
 + \mathcal O(z^{2r+1}).
 \label{Padexp}
\ee
The polynomial $P_n$ is explicitly known\footnote{We have normalized the polynomial to be monic.}  (\cite{Perron}, p. 433) 
\be
\label{Padexpn}
P_r(z) = \sum_{k=0}^r \frac{(2r-k)!(-z)^k}{ k!(r-k)!} .
\ee
The location of the zeros $\mathfrak Z_r:= \{a_j, j=1,\dots, r\} $, of $P_r (z)$   (plotted in Fig. \ref{plotzeros} by way of example) is  known to belong to the
annular sector \cite{SaffVarga78}
\be
\label{estSaffVarga}
2r\mu<|a_j|< 2r+\frac 43,\ \ \mu>0,\  \mu{\rm e}^{1+\mu}=1 \qquad (\mu \simeq 0.278465...)\\
\label{minrea}
|\arg(a_j)| \leq \cos^{-1} (1/r)\qquad 
\Re (a_j) >2 \mu>\frac 1 2.
\ee

\begin{figure}
\begin{minipage}{0.45\textwidth}
\begin{center}
\begin{tikzpicture}[scale=2]
\draw [fill=  white!47!black!87!blue, fill opacity = 0.3](0, 1+2/3/70)  arc (90:-90:1+2/3/70);
\draw [fill=  white] (0, 2*0.278465)  arc (90:-90:2*0.278465);
\draw [step=0.1, gray!30!white, line width = 0.1pt] (-1.1,-1.1) grid (1.1,1.1);
\draw [step=0.5, gray!70!white, line width = 0.1pt] (-1,-1) grid (1,1);
\node [left] at (0,2*0.278465) {$\mu$};
\node[left] at (0,1) {$1 + \frac {2}{3n}$};
\draw (0, 1+2/3/70)  arc (90:-90:1+2/3/70);
\draw (0, 2*0.278465)  arc (90:-90:2*0.278465);

\foreach \x in {(0.91750455562578326123616203256443595008539858632403e-1, -.95059712318798989974400663374186400811726047057713), (0.91750455562578326123616203256443595008539858632403e-1, .95059712318798989974400663374186400811726047057713), 
(.15636043849418927029270556636258152717661314088658, -.90592907127440083777403351070910049947445408549837),
(.15636043849418927029270556636258152717661314088658, .90592907127440083777403351070910049947445408549837),
(.20666143173388094445533457100166352325077552939808, -.86786047850242823039499899992773393773405143493820), 
(.20666143173388094445533457100166352325077552939808, .86786047850242823039499899992773393773405143493820), 
(.24916176412430588387529564174695685171438927403850, -.83309190540314343181238484675336686878108502352206),
(.24916176412430588387529564174695685171438927403850, .83309190540314343181238484675336686878108502352206), 
(.28639682239941638573435987430929281913585269386261, -.80038709558929352696583146542881705868869526462084),
(.28639682239941638573435987430929281913585269386261, .80038709558929352696583146542881705868869526462084), 
(.31969433944848107517123082121102178545322927833507, -.76912017663698672235329968598305636263445792978385), 
(.31969433944848107517123082121102178545322927833507, .76912017663698672235329968598305636263445792978385), 
(.34986168104015313527506014252128990345852527302847, -.73892197336366195158683882891019318783784390325224), 
(.34986168104015313527506014252128990345852527302847, .73892197336366195158683882891019318783784390325224), 
(.37743701361211405865344877381896307515257403282318, -.70955313855915322491868247297951979666111226701088), 
(.37743701361211405865344877381896307515257403282318, .70955313855915322491868247297951979666111226701088), 
(.40280218429951585797220069377881353854614800515713, -.68084811133773424905095718285476562370572327674268), 
(.40280218429951585797220069377881353854614800515713, .68084811133773424905095718285476562370572327674268), 
(.42624066753977011797461757129495593885230883113864, -.65268681230615665892374881521604944820608168750431), 
(.42624066753977011797461757129495593885230883113864, .65268681230615665892374881521604944820608168750431), 
(.44797023612907103409459748266442523341752975998112, -.62497892960027051793848097425597571658115785576008), 
(.44797023612907103409459748266442523341752975998112, .62497892960027051793848097425597571658115785576008), 
(.46816271808232802320261686111225527449937998482543, -.59765455554900590921048171970615798793996074376135), 
(.46816271808232802320261686111225527449937998482543, .59765455554900590921048171970615798793996074376135), 
(.48695661266170900183578597029293182260887143655012, -.57065829148879608461886346116949776687200517618235), 
(.48695661266170900183578597029293182260887143655012, .57065829148879608461886346116949776687200517618235), 
(.50446550558568757443042848238131934281730394873655, -.54394536837671133870595358693829233203924090660468), 
(.50446550558568757443042848238131934281730394873655, .54394536837671133870595358693829233203924090660468), 
(.52078388627887291268340802550280269994325739751138, -.51747899947918986728982155814902438770308325082085), 
(.52078388627887291268340802550280269994325739751138, .51747899947918986728982155814902438770308325082085), 
(.53599129073886564307527376782993927193805034498981, -.49122851801587023219873435082917515867046467665197), 
(.53599129073886564307527376782993927193805034498981, .49122851801587023219873435082917515867046467665197), 
(.55015532716141318697588432906942177501217816653193, -.46516803262295216956336475816217377993463927752757), 
(.55015532716141318697588432906942177501217816653193, .46516803262295216956336475816217377993463927752757), 
(.56333393366393494546203314172678961209726718021782, -.43927543469457241164404382660772360533885128268155), 
(.56333393366393494546203314172678961209726718021782, .43927543469457241164404382660772360533885128268155), 
(.57557709451227909505402933734272789773834333293956, -.41353165102558359084634378676239385102049523101106), 
(.57557709451227909505402933734272789773834333293956, .41353165102558359084634378676239385102049523101106), 
(.58692816583625636544985864531845900595309269996993, -.38792007129885087610207977219466513654971190902375), 
(.58692816583625636544985864531845900595309269996993, .38792007129885087610207977219466513654971190902375), 
(.59742491406686996755231993860036688850492674780209, -.36242610264363562777347286908985952428619805070133), 
(.59742491406686996755231993860036688850492674780209, .36242610264363562777347286908985952428619805070133), 
(.60710033924351197251410186692243840416329394631076, -.33703681813778505928399840423594409916398231066738), 
(.60710033924351197251410186692243840416329394631076, .33703681813778505928399840423594409916398231066738), 
(.61598333460178559436149499372462675653706979725442, -.31174067581764228921429911901800298073048414517044), 
(.61598333460178559436149499372462675653706979725442, .31174067581764228921429911901800298073048414517044), 
(.62409921971090373552203058809300070068109839929977, -.28652729131340126639612850082426695623199399611953), 
(.62409921971090373552203058809300070068109839929977, .28652729131340126639612850082426695623199399611953), 
(.63147017459270349815957377589878164608561412154674, -.26138725174647590900331734769760331609306219724016), 
(.63147017459270349815957377589878164608561412154674, .26138725174647590900331734769760331609306219724016), 
(.63811559528823537180145891335864989233683597175651, -.23631196169617636230327402102558578482574874075649), 
(.63811559528823537180145891335864989233683597175651, .23631196169617636230327402102558578482574874075649), 
(.64405238631979262024309868244419786493697363447179, -.21129351430225775625761424185826858494794602417510), 
(.64405238631979262024309868244419786493697363447179, .21129351430225775625761424185826858494794602417510), 
(.64929520182984117322704333433220260812395124754086, -.18632458220153098470724698083637208077498889473021), 
(.64929520182984117322704333433220260812395124754086, .18632458220153098470724698083637208077498889473021), 
(.65385664445024943513787304761069546300437638659731, -.16139832419258658483560157729080030113626943535195), 
(.65385664445024943513787304761069546300437638659731, .16139832419258658483560157729080030113626943535195), 
(.65774742891068909500975956162722318884436624047897, -.13650830439853672599948896369056156328755629404780), 
(.65774742891068909500975956162722318884436624047897, .13650830439853672599948896369056156328755629404780), 
(.66097651581981851390276020834799958307053677115609, -.11164842136537267308553321619632466859095646546300), 
(.66097651581981851390276020834799958307053677115609, .11164842136537267308553321619632466859095646546300), 
(.66355121984349803895452972614904954964152706709986, -0.86812845006331906742704722014246496963408896988550e-1), 
(.66355121984349803895452972614904954964152706709986, 0.86812845006331906742704722014246496963408896988550e-1), 
(.66547729554092494774408147375617983461875430847841, -0.61995959697166347328190374209340755147821383913445e-1), 
(.66547729554092494774408147375617983461875430847841, 0.61995959697166347328190374209340755147821383913445e-1), 
(.66675900334301156426585550183775545567790080011662, -0.37192312070051990937270687283390805497891875056734e-1), 
(.66675900334301156426585550183775545567790080011662, 0.37192312070051990937270687283390805497891875056734e-1), 
(.66739915753334163381223248475377766999854439053582, -0.12396562289696211562024303683242139929405358938469e-1), 
(.66739915753334163381223248475377766999854439053582, 0.12396562289696211562024303683242139929405358938469e-1)}
{
\draw[fill] \x circle [radius=0.3pt];
};

\draw (-1.1,0)--(1.1,0);
\draw (0,-1.1)--(0,1.1);

\end{tikzpicture}
\end{center}
\caption{The zeroes of $P_n(2nz)$ for $n=70$. }
\label{plotzeros}
\end{minipage}
\hfill
\begin{minipage}{0.45\textwidth}

\begin{tikzpicture}[scale=2]
\begin{scope}[xscale=-1]
\draw [fill=  white!47!black!87!red, fill opacity = 0.3](0, 1+2/3/70)  arc (90:-90:1+2/3/70);
\draw [fill=  white] (0, 2*0.278465)  arc (90:-90:2*0.278465);
\draw [step=0.1, gray!30!white, line width = 0.1pt] (-1.1,-1.1) grid (1.1,1.1);
\draw [step=0.5, gray!70!white, line width = 0.1pt] (-1,-1) grid (1,1);
\node [below left] at (0,2*0.278465) {$2n\mu$};
\node[above left] at (0,1) {$2n + \frac {4}{3}$};
\draw (0, 1+2/3/70)  arc (90:-90:1+2/3/70);
\draw (0, 2*0.278465)  arc (90:-90:2*0.278465);
\foreach \x in {(0.91750455562578326123616203256443595008539858632403e-1, -.95059712318798989974400663374186400811726047057713), (0.91750455562578326123616203256443595008539858632403e-1, .95059712318798989974400663374186400811726047057713), 
(.15636043849418927029270556636258152717661314088658, -.90592907127440083777403351070910049947445408549837),
(.15636043849418927029270556636258152717661314088658, .90592907127440083777403351070910049947445408549837),
(.20666143173388094445533457100166352325077552939808, -.86786047850242823039499899992773393773405143493820), 
(.20666143173388094445533457100166352325077552939808, .86786047850242823039499899992773393773405143493820), 
(.24916176412430588387529564174695685171438927403850, -.83309190540314343181238484675336686878108502352206),
(.24916176412430588387529564174695685171438927403850, .83309190540314343181238484675336686878108502352206), 
(.28639682239941638573435987430929281913585269386261, -.80038709558929352696583146542881705868869526462084),
(.28639682239941638573435987430929281913585269386261, .80038709558929352696583146542881705868869526462084), 
(.31969433944848107517123082121102178545322927833507, -.76912017663698672235329968598305636263445792978385), 
(.31969433944848107517123082121102178545322927833507, .76912017663698672235329968598305636263445792978385), 
(.34986168104015313527506014252128990345852527302847, -.73892197336366195158683882891019318783784390325224), 
(.34986168104015313527506014252128990345852527302847, .73892197336366195158683882891019318783784390325224), 
(.37743701361211405865344877381896307515257403282318, -.70955313855915322491868247297951979666111226701088), 
(.37743701361211405865344877381896307515257403282318, .70955313855915322491868247297951979666111226701088), 
(.40280218429951585797220069377881353854614800515713, -.68084811133773424905095718285476562370572327674268), 
(.40280218429951585797220069377881353854614800515713, .68084811133773424905095718285476562370572327674268), 
(.42624066753977011797461757129495593885230883113864, -.65268681230615665892374881521604944820608168750431), 
(.42624066753977011797461757129495593885230883113864, .65268681230615665892374881521604944820608168750431), 
(.44797023612907103409459748266442523341752975998112, -.62497892960027051793848097425597571658115785576008), 
(.44797023612907103409459748266442523341752975998112, .62497892960027051793848097425597571658115785576008), 
(.46816271808232802320261686111225527449937998482543, -.59765455554900590921048171970615798793996074376135), 
(.46816271808232802320261686111225527449937998482543, .59765455554900590921048171970615798793996074376135), 
(.48695661266170900183578597029293182260887143655012, -.57065829148879608461886346116949776687200517618235), 
(.48695661266170900183578597029293182260887143655012, .57065829148879608461886346116949776687200517618235), 
(.50446550558568757443042848238131934281730394873655, -.54394536837671133870595358693829233203924090660468), 
(.50446550558568757443042848238131934281730394873655, .54394536837671133870595358693829233203924090660468), 
(.52078388627887291268340802550280269994325739751138, -.51747899947918986728982155814902438770308325082085), 
(.52078388627887291268340802550280269994325739751138, .51747899947918986728982155814902438770308325082085), 
(.53599129073886564307527376782993927193805034498981, -.49122851801587023219873435082917515867046467665197), 
(.53599129073886564307527376782993927193805034498981, .49122851801587023219873435082917515867046467665197), 
(.55015532716141318697588432906942177501217816653193, -.46516803262295216956336475816217377993463927752757), 
(.55015532716141318697588432906942177501217816653193, .46516803262295216956336475816217377993463927752757), 
(.56333393366393494546203314172678961209726718021782, -.43927543469457241164404382660772360533885128268155), 
(.56333393366393494546203314172678961209726718021782, .43927543469457241164404382660772360533885128268155), 
(.57557709451227909505402933734272789773834333293956, -.41353165102558359084634378676239385102049523101106), 
(.57557709451227909505402933734272789773834333293956, .41353165102558359084634378676239385102049523101106), 
(.58692816583625636544985864531845900595309269996993, -.38792007129885087610207977219466513654971190902375), 
(.58692816583625636544985864531845900595309269996993, .38792007129885087610207977219466513654971190902375), 
(.59742491406686996755231993860036688850492674780209, -.36242610264363562777347286908985952428619805070133), 
(.59742491406686996755231993860036688850492674780209, .36242610264363562777347286908985952428619805070133), 
(.60710033924351197251410186692243840416329394631076, -.33703681813778505928399840423594409916398231066738), 
(.60710033924351197251410186692243840416329394631076, .33703681813778505928399840423594409916398231066738), 
(.61598333460178559436149499372462675653706979725442, -.31174067581764228921429911901800298073048414517044), 
(.61598333460178559436149499372462675653706979725442, .31174067581764228921429911901800298073048414517044), 
(.62409921971090373552203058809300070068109839929977, -.28652729131340126639612850082426695623199399611953), 
(.62409921971090373552203058809300070068109839929977, .28652729131340126639612850082426695623199399611953), 
(.63147017459270349815957377589878164608561412154674, -.26138725174647590900331734769760331609306219724016), 
(.63147017459270349815957377589878164608561412154674, .26138725174647590900331734769760331609306219724016), 
(.63811559528823537180145891335864989233683597175651, -.23631196169617636230327402102558578482574874075649), 
(.63811559528823537180145891335864989233683597175651, .23631196169617636230327402102558578482574874075649), 
(.64405238631979262024309868244419786493697363447179, -.21129351430225775625761424185826858494794602417510), 
(.64405238631979262024309868244419786493697363447179, .21129351430225775625761424185826858494794602417510), 
(.64929520182984117322704333433220260812395124754086, -.18632458220153098470724698083637208077498889473021), 
(.64929520182984117322704333433220260812395124754086, .18632458220153098470724698083637208077498889473021), 
(.65385664445024943513787304761069546300437638659731, -.16139832419258658483560157729080030113626943535195), 
(.65385664445024943513787304761069546300437638659731, .16139832419258658483560157729080030113626943535195), 
(.65774742891068909500975956162722318884436624047897, -.13650830439853672599948896369056156328755629404780), 
(.65774742891068909500975956162722318884436624047897, .13650830439853672599948896369056156328755629404780), 
(.66097651581981851390276020834799958307053677115609, -.11164842136537267308553321619632466859095646546300), 
(.66097651581981851390276020834799958307053677115609, .11164842136537267308553321619632466859095646546300), 
(.66355121984349803895452972614904954964152706709986, -0.86812845006331906742704722014246496963408896988550e-1), 
(.66355121984349803895452972614904954964152706709986, 0.86812845006331906742704722014246496963408896988550e-1), 
(.66547729554092494774408147375617983461875430847841, -0.61995959697166347328190374209340755147821383913445e-1), 
(.66547729554092494774408147375617983461875430847841, 0.61995959697166347328190374209340755147821383913445e-1), 
(.66675900334301156426585550183775545567790080011662, -0.37192312070051990937270687283390805497891875056734e-1), 
(.66675900334301156426585550183775545567790080011662, 0.37192312070051990937270687283390805497891875056734e-1), 
(.66739915753334163381223248475377766999854439053582, -0.12396562289696211562024303683242139929405358938469e-1), 
(.66739915753334163381223248475377766999854439053582, 0.12396562289696211562024303683242139929405358938469e-1)}
{
\draw[fill, black!50!red] \x circle [radius=0.3pt];
};
\end{scope}
\draw (0,0)-- (75:1.2);

\draw [<->] (75:1.2) arc (75:90:1.2);
\node at  (82:1) {$\theta_0$};
\draw (-1.1,0)--(1.1,0);
\draw (0,-1.1)--(0,1.1);

\end{tikzpicture}
\caption{The poles of the Pad\'e\ approximation are within the shaded sectorial annulus.}\label{sectestim}
\end{minipage}
\end{figure}

\paragraph{Estimate for the remainder.}

The remainder of the approximation is also known exactly 
\bea
\label{remainder}
{\rm e}^{-z} - \frac {P_r(z)}{P_r(-z)}  =\frac {(-1)^{r+1} z^{2r+1}}{r!P_r(-z) } \int_0^ 1
 {\rm e}^{-t z} (1-t)^r t^r \d t.
\eea
The estimate \eqref{estSaffVarga} on the position of zeroes implies that for $\Re (z)\geq 0$, the minimum distance from the poles $-\mathfrak Z_r $ for $|\arg (z)| \leq \frac \pi 2 - \theta_0$  for some small fixed $\theta_0$, is (see Fig. \ref{sectestim})
\be
{\rm dist} (z,- \mathfrak Z_r) \geq
 \frac {1}{\sqrt{ (|z|-2r\mu \cos \theta_0 )^2 + (2r\mu\sin\theta_0)^2 } }
\geq  \frac {2}{ |z| + 2r\mu \sin\theta_0 }.
\ee
Then we can estimate 
\be
\le|{\rm e}^{-z} - \frac {P_r(z)}{P_r(-z)} \ri| \leq  \frac {  2 |z|^{r}}{r! (\sin\theta_0)^r(2r\mu\sin\theta_0 + |z|)^r } \int_0^{\Re  z}\!\!\!\! {\rm e}^{-t }  t^r \d t  
 \leq   \frac {  2 |z|^{r}}{ (\sin\theta_0)^r( 2n\mu \sin\theta_0+ |z|)^r }.
\label{EstimPade}
\ee

Let $N\in \N$ be fixed. We want to obtain an approximation of ${\rm e}^{-2t \l^{\frac {2N+1}2}}$ in a suitable sector; of course we will use \eqref{Padexp} replacing $ z\mapsto 2t \l^{\frac {2N+1}2}$.

\noindent {\bf Proof of Thm. \ref{ThmP1N}$_{[1]}$.}
From \eqref{EstimPade} we  have (we set $t = t_{2N+1}$, $n= r(2N+1)$ for brevity)
\be
\frac {P_{r}(2 t  \l^{\frac{2N+1}2} )}{P_{r}(- 2 t \l ^{\frac{2N+1}2})}- {\rm e}^{-t \l^{\frac{2N+1}2}} =\mathcal O\le(
\frac {   |t|^r |\l|^{\frac {r(2N+1)}2 }}{ (\sin\theta_0)^r(( 2n\mu\sin(\theta_0))^r +  |t|^r |\l|^{\frac {r(2N+1)}2}) }
\ri),
\nonumber \\
\arg \l \in \mathcal J_0(k,t) := \le(-\frac \pi{2N+1}, \frac {\pi}{2N+1}\ri) + \frac { 4\pi k_0-2\arg(t)}{2N+1}.
\ee
The estimate above shows that inside the growing disk  $\le |t\l^\frac {2N+1}{2}\ri|\leq K   r^\frac 1 2$  (with $K= (4N+2)(\sin^2  \theta_0)$) the expression is bounded as follows:
\be
\label{est3}
\le| \frac {P_{r}(2 t  \l^{\frac{2N+1}2} )}{P_{r}(- 2 t \l ^{\frac{2N+1}2})}- {\rm e}^{-t \l^{\frac{2N+1}2}}\ri|  \leq  \le\{ 
\begin{array}{cc}
 { r^{-\frac r 2}} & |\l| \leq  K ^\frac {2}{2N+1} |t|^{\frac {-2}{2N+1}} r ^{\frac {1}{2N+1}}, \ \arg(\l) \in \mathcal J_0\\
1 &  |\l| \geq   K ^\frac {2}{2N+1}|t|^{\frac {-2}{2N+1}} r ^{\frac {1}{2N+1}}, \ \arg(\l) \in \mathcal J_0.
\end{array}
\ri.
\ee
 
Let $\G_n(\l; t, x)$ ($n=r(2N+1)$)  be the solution of the RHP \eqref{RHPgamma} with ${\mathbf d}_n=\frac {P_{r}( t  \l^{\frac{2N+1}2} )}{P_{r}(- t \l ^{\frac{2N+1}2})}$ and 
let $\G(\l; t,x)$ be the solution discussed in Proposition \ref{propP1toAiry}.
 Similar estimates hold for $\frac {P_{r}(-2 t  \l^{\frac{2N+1}2} )}{P_{r}( 2 t \l ^{\frac{2N+1}2})}- {\rm e}^{t \l^{\frac{2N+1}2}}$ in the sectors $\mathcal J_\pm$ \eqref{rangeP1N}.  We now choose $k_+, k_0,k_-$ and the corresponding ways $\varpi_{0,\pm}$; the tau function  $\tau_n(z;\vec \l,\vec \mu)$  is then given by \eqref{ZnKcont} according to Theorem \ref{main2bis}$_{[1]}$. The relationship between the positioning of the jump rays $\varpi_{0,\pm}$ and the integers $k_{0,\pm}$ follows from the formula \eqref{GnK} and careful inspection. 

It only remains to show that the solution of the Riemann--Hilbert problem \ref{RHPgamma} converges to the solution of the Painlev\'e\ auxiliary Riemann--Hilbert problem \ref{RHPP1N}, which we now accomplish. 

Following the same idea as in Proposition \ref{propP1toAiry}, let $\mathcal E=\mathcal E(\l; t,x,n)$ be the solution of the Riemann--Hilbert problem with jumps on $\varpi_{0,\pm 1}$ of the form 
\be
\mathcal E_+ = \mathcal E_- \G 
\bigg(\1 + {\rm e}^{-\frac 23  \l ^\frac 32 - x \sqrt\l}
\overbrace{ \bigg(\frac {P_{r}( t  \l^{\frac{2N+1}2} )}{P_{r}(- t \l ^{\frac{2N+1}2})}- {\rm e}^{-t \l^{\frac{2N+1}2}}\bigg)}^{:= \mathfrak F_0(\l)}
 \s_+ \bigg)
 \G^{-1} \ ,\ \ \l \in \varpi_0
\label{jumpE1}\\
\mathcal  E_+ =  \mathcal E_- \G 
\bigg(\1 + {\rm e}^{\frac 23  \l ^\frac 32 + x \sqrt\l} 
\underbrace{\bigg(\frac {P_{r}( -t  \l^{\frac{2N+1}2} )}{P_{r}( t \l ^{\frac{2N+1}2})}- {\rm e}^{t \l^{\frac{2N+1}2}}\bigg)}_{:= \mathfrak F_\pm (\l)}
\s_- \bigg)\G^{-1} \ ,\ \ \l \in \varpi_\pm
 \label{jumpE2}\\
 \mathcal E(\l;t,x,n) = \1 + \frac 1\l \mathcal E_1(t,x,n) + \mathcal O(\l^{-2}),\ \ \l\to\infty.
\ee

Consider  \eqref{jumpE1}, with similar considerations for \eqref{jumpE2}: given the estimate \eqref{est3} we have that, uniformly for $x$ in compact sets, 
\be
\le|{\rm e}^{-\frac 23  \l ^\frac 32 - x \sqrt\l} \mathfrak F_0(\l) \ri| = \le\{
\begin{array}{cc}
\mathcal O(r^{-\frac r 2})  &  |\l| \leq  |t|^{\frac {-2}{2N+1}} r ^{\frac {1}{2N+1}}, \ \l\in \varpi_0
\\
 \mathcal O\le(
 \exp\le(
 -C |t|^{\frac {-3}{2N+1}} r ^{\frac {3}{4N+2}}
 \ri)
 \ri)  &  |\l| \geq  |t|^{\frac {-2}{2N+1}} r ^{\frac {1}{2N+1}}, \ \ \l\in \varpi_0.
\end{array}
\ri.
\label{est4}
\ee
According to Proposition \ref{propP1toAiry} for $|t|$ sufficiently small and $x$ in a compact set, the  function $\G(\l;t, x)$ remains uniformly bounded and therefore it is easy to see, using \eqref{est4}, that  the jump matrices \eqref{jumpE1}, \eqref{jumpE2} converge to the identity in all $L^p$ norms ($1 \leq p\leq \infty$)  as $r\to \infty$, and hence so does $\mathcal E$; the rate of convergence is the same as in \eqref{est4} and it is faster than any inverse power of $r$ and thus on $n$. Note that the angle of the ray $\varpi_0$ ranges in a sector where there are zeroes of the numerator in $\mathfrak F_{0}(\l)$ and since the zeroes and poles of the numerator/denominator are contained in disjoint sectors, the function $\mathfrak F_0(\l)$ is bounded and analytic along the ray $\varpi_0$. Similarly for the other two rays.

Then, by the same token used in Proposition \ref{propP1toAiry} we must have
\be
\G_n(\l;t,x) = \bigg(\1 - \mathcal E_1(t,x,n)_{21}\s_-\bigg) \mathcal E(\l;t,x,n) \G(\l;t,x)
\ee
and we conclude that the tau function for the problem $\G_n$ (i.e. $Z_n$) converges to the tau function of $\G$. 
\QED

\appendix
\renewcommand{\theequation}{\Alph{section}.\arabic{equation}}

\section{Proof of Thm. \ref{taudressing}}
\label{proofSchles}
Let $\G_n = \G_n(\l;\t,\vec \l, \vec \mu)$ denote the solution of the Riemann--Hilbert problem with the jump matrices $M_n(\l;\t,\vec \l, \vec \mu)$. 
It can be written as $\G_n(\l) = R(\l) \G_0(\l) D(\l)$ where $R(\l)$ is a suitable {\it rational} matrix and $D(\l)$ as in \eqref{defD}. Indeed the matrix ratio 
\bea
&&R(\l):= \G_n (\l)D(\l)^{-1} \G_0(\l) ^{-1}\label{316}
\eea
is seen to have no jumps on $\Sigma$. It clearly has at most simple poles at $\l=\l_k, \mu_k$ and decays  algebraically at infinity. By Liouville's theorem $R(\l)$ is a rational function.

We now seek a set of characterizing conditions for the matrix $R$ as a solution of a given Riemann--Hilbert problem. To this end let $r > 0$ be sufficiently small so that all the disks below are disjoint and define:

\noindent
\begin{minipage}{0.4\textwidth}
\begin{tikzpicture}
\draw [fill= white!94!black!80!red, fill opacity = 0.3] (0,0) circle [radius =2];
\draw [fill= white] (20:1) circle [radius =0.4] node {$\mathbb D_1$};
\draw [fill= white] (70:1.2) circle [radius =0.4] node {$\mathbb D_2$};
\draw [fill= white] (-70:1) circle [radius =0.4] node {$\mathbb D_k$};
\draw [fill= white] (-120:1.4) circle [radius =0.4] node {$\mathbb D_n$};
\node at (0:3) {$\mathbb D_{\infty}$};
\node at (180:1) {$\mathbb D_{-}$};
\end{tikzpicture}
\end{minipage}\hspace{0.031\textwidth}
\begin{minipage}{0.56\textwidth}
 \bea
&\&\mathbb D_+:=\mathbb D_\infty\cup  \bigsqcup_{k=1}^{n} \mathbb D_k \ ,\nonumber\\
 &\&\mathbb D_k := \{|\l-\l_k| < r\}\ , k\leq n_2,   \nonumber\\
&\&  \mathbb D_{k+n_2} := \{|\l-\mu_k| < r\}\ , k\leq n_1\nonumber\\
&\& 
\mathbb D_\infty := \le\{\le|\frac 1\l\ri|>  r\ri \}\nonumber\\
&\&\mathbb D_- := \C \setminus \overline{\mathbb D_+}.
\eea
\end{minipage}
\begin{problem}\label{RHPR}
Find a $2\times 2$ piecewise analytic function $\RR(\l)$ on  $\mathbb D_+$ and $\mathbb D_-$, admitting continuous boundary values and satisfying the following conditions
\be
\label{RHPRR} 
\RR_+(\l) = \RR_-(\l) \J(\l)\ ,\qquad
\RR(\l) = \1 + \mathcal O(\l^{-1}) \ ,\ \ \l\to \infty
\ee 
where $\J(\l)$ is the matrix that on each component of $\pa \mathbb D_+$ restricts to 
\bea
\label{defJ} 
\J(\l) =  \le\{
\begin{array}{ll}
\ds  J_k(\l) = \G_0(\l) (\l_k-\l)^{E_{22}} \ ,& |\l-\l_k| = r,\ k\leq n_2
\\[8pt]
\ds  J_{n_2+k}(\l) = \G_0(\l) (\mu_k-\l)^{E_{11}} \ ,& |\l-\mu_k| = r,\ k\leq n_1
\\[8pt]
\ds
 J_\infty(\l) = \G_0(\l)D(\l) \frac {\1- i\s_1} {\sqrt{2}}  \l^{\frac {\s_3}4} & |\l|^{-1} = r.
 \end{array}\ri.
 \eea
\end{problem}
 We shall use interchangeably $\RR_\pm(\l)$ for the boundary values or the restriction of the solution $\RR(\l)$ to $\mathbb D_\pm$, respectively. 
We prove the following:
\bp
	The matrix $R(\l)$ in \eqref{316}  is a rational matrix--valued function with simple poles at $\l=\l_k, \mu_k$. Restricted to $\mathbb D_-$ it coincides with $\RR_-(\l)$ in the Riemann--Hilbert problem \ref{RHPR}, up to a left constant  multiplier of the form $\1 - ia^{(n)}\s_-$. In particular, the matrix $\RR_-(\l)$ extends to a rational function of $\l$.
\ep
\noindent {\bf Proof.} 
We already proved that $R(\l)$ is a rational function. Moreover near $\l_k, \mu_k $ we have, from \eqref{316},  
\be
R(\l) =\mathcal O(1) (\l-\l_k)^{-E_{22}} \G_0^{-1}(\l),\qquad 
R(\l)= \mathcal O(1) (\l-\mu_k)^{-E_{11}}\G_0^{-1}(\l)
\ee
and thus $R(\l)$   must have simple poles at $\l_k,\mu_k$ (here $\mathcal O(1)$ stands for a locally analytic matrix function, with analytic inverse).
Now in order to establish the connection between $\RR$ and $R$ we have to study the behaviour at $\l=\infty$. We multiply both sides of $\G_n = R \G_0 D$ ($D$ defined in \eqref{defD}) as follows
\bea
\label{Rinfty}
\G_n(\l)  \le(\l^{-\frac {\s_3}4} \frac {\1 + i\s_1}{\sqrt{2}} \ri)^{-1}= R(\l)\underbrace{ \G_0(\l) D (\l) \frac {1-i\s_1}{\sqrt{2}} \l^{\frac {\s_3}4}}_{J_\infty(\l)}.
\eea
Because of the asymptotic behavior \eqref{Gninfty} the left side admits a regular expansion at $\l=\infty$ with leading coefficient of the form $C_0 = \1 + i a^{(n)} \s_-$.

Hence we define $\RR(\l)$ as 
\be\label{defRR}
\RR(\l) = \le\{
\begin{array}{ll}
\le(\1 - i a^{(n)}\s_- \ri)\G_n(\l)  D^{-1} (\l) (\l-\l_k)^{E_{22}}& \l\in \mathbb D_k, \ k=1,\dots, n_2,\\[3pt]
\le(\1 - i a^{(n)}\s_- \ri)\G_n(\l)  D^{-1} (\l) (\l-\mu_k)^{E_{11}}& \l\in \mathbb D_{n_2+k}, \ k=1,\dots, n_1,\\[3pt]
\le(\1 - i a^{(n)}\s_- \ri)\G_n(\l)  \frac {1-i\s_1}{\sqrt{2}} \l^{\frac {\s_3}4} &
\l\in \mathbb D_\infty , \\[3pt]
\le(\1 - i a^{(n)}\s_- \ri)\G_n(\l) D^{-1}(\l) \G_0^{-1}(\l) & \l \in \mathbb D_-.
\end{array}
\ri. 
\ee
It is easy to check that $\RR$, defined in this way, satisfies the jump conditions in \eqref{RHPRR} and, moreover, also the asymptotic condition at infinity. The identification (up to a normalization constant) between $\RR_-$ and $R$ is read off directly from the last line of \eqref{defRR}. \QED

The jump matrix $\J(\l)$ admits (formal) meromorphic extension in the interior of $\mathbb D$ (by construction) and moreover the total index of $\det \J(\l)$ around $\pa \mathbb D$ is zero. 
Under these conditions, the analysis in Appendix B of \cite{BertolaCafasso5} applies. We briefly remind it with notation adapted to the current use.

Let $\H_\pm$ denote the vector space of  (formally) analytic valued row-vectors in $\mathbb D_\pm$ (respectively) and $C_\pm$ the Cauchy projection operator; consider the following two (finite dimensional, see B.14 in loc. cit.) subspaces of $\H_-$
\be
V:= C_-[\H_+ \J^{-1}(\l)] \ ,\qquad W:= C_-[\H_+ \J(\l)]
\ee
\bp[Prop. B.3 in \cite{BertolaCafasso5}]
The solution of the  Riemann--Hilbert Problem \ref{RHPR} exists, and is unique, if and only if  the linear map 
\be
\label{Gmap}
\begin{array}{cccc}
\mathcal G:& V&\longrightarrow& W\\
& v & \longmapsto &C_-[v \J].
\end{array}
\ee
is invertible. In this case, the inverse is 
\be
\begin{array}{cccl}
\mathcal G^{-1}:&W& \longrightarrow & V\\
&w&\longmapsto & C_-\le[w \J^{-1} \RR^{-1}\ri]\RR.
\end{array}
\label{inverseG}
\ee
\ep
\br
Even if $R$ and $\RR$ differ by the multiplication of a {\it constant} left multiplier of the form $\1 + \star \s_-$, the expression of the inverse \eqref{inverseG} is unaffected by such multiplier.
\er

Now, following  \cite{BertolaCafasso5}, we choose properly two bases for $V$ and $W$ in such a way that the determinant $\mathbb G$ of $\mathcal G$ gives the variation of the one form $\Omega$ as in \eqref{varOmega}.

The jump on the large circle given by $J_\infty$ \eqref{defJ}  has an asymptotic expansion in {\it integer} powers of $\l$ of the form (the $\star$ symbol means a constant of no interest to us, which turns out to be $-\frac {x^2}4$) 
\bea
 {\bf n=2k\equiv 0 \ {\rm mod}\  2} 
 && J_\infty(\l)  =(-1)^{n_1}  \l^{k}\le( \1 + \star \s_- + \mathcal O(\l^{-1})\ri)\cr
 &&=: G_{\infty}(\l) \l^{k\1},
 \label{Jinfty0}
\\
{\bf n=2k+1\equiv 1 \ {\rm mod}\  2} 
&& 
J_\infty(\l)  =(-1)^{n_1}  \l^{k} \le(i\l \s_- -i\s_+   + \star \1 + \mathcal O(\l^{-1})\ri) =\nonumber\\[3pt]
&& =(-1)^{n_1} \le(\s_2 + \star E_{22} + \mathcal O(\l^{-1})\ri) \l^{(k+1) E_{11} + k E_{22}}\nonumber \\[3pt]
&& 
=:G_{\infty}(\l) \l^{(k+1) E_{11} + k E_{22}},
\label{Jinfty}
\eea
where $G_\infty(\l)$ (as in \eqref{Ginfty}) has been introduced for convenience.
Following \cite{BertolaCafasso5}, Appendix B,  we choose the bases
\bea
V  = \bigoplus_{k=1}^{n_2+n_1}\C \{v_{k}\}; &\& \qquad
\begin{array}{ll}
v_{j} = 
 C_-\big[ {\mathbf e_2^{\mathrm T}}\J^{-1}(\l) \big]  = \frac{\mathbf e_2^{\mathrm T}\G_0^{-1}(\l_j)}{\l-\l_j}\ , &1\leq j\leq n_2\
 \\ [3pt]
 v_{j+n_2} = 
 C_-\big[ {\mathbf e_1^{\mathrm T}}\J^{-1}(\l) \big]  = \frac{\mathbf e_1^{\mathrm T}\G_0^{-1}(\mu_j)}{\l-\mu_j}\ , &1\leq  j\leq n_1,
 \end{array}
 \label{Vbasis}
\\
W =\bigoplus_{\ell=1}^{n_2+n_1} \C \{w_{\ell}\}; &\& \qquad
w_{n-\ell+1} = 
 {{\bf e}_{(\ell-1 {\rm mod} 2)+1}^{\mathrm T}}\l^{\lfloor (\ell-1)/2\rfloor} , 
 \qquad 1\leq \ell \leq n.
 \label{Wbasis}
\eea
The basis of $W= C_-[\H_+ \J]$ is obtained by noticing the vector space is the same as $C_-[\H_+ G_\infty^{-1} \J]$ (because $G_\infty$ is (formally) analytic at $\l=\infty$) and hence it is the same as $C_-[\H_+  \l^{\lfloor (n+1)/2\rfloor E_{11} + \lfloor n/2\rfloor E_{22}}]$.
The matrix $\mathbb G_{k,\ell}$ representing $\mathcal G$ \eqref{Gmap} for $k\leq n_2$ is then given by a direct computation as 
\bea
\mathbb G_{k,\ell} &\& = \res{\lambda=\infty} \res{\z=\l_k} \frac{\mathbf e_2^{\mathrm T}\G_0^{-1}(\l_k)}{(\z-\l_k)(\l-\z)} G_{\infty}(\l) \l^{\lfloor \frac {n+1}2\rfloor E_{11} +\lfloor \frac {n}2\rfloor E_{22}}
\frac {{\bf e}_{(\ell - 1 {\rm mod}\  2 )+1}}{\l^{\lfloor (\ell-1)/2\rfloor+1}} 
\\
&\& = \res{\lambda=\infty}  \frac{\mathbf e_2^{\mathrm T}\G_0^{-1}(\l_k)}{(\l-\l_k)} G_{\infty}(\l) \l^{\lfloor \frac {n+1}2\rfloor E_{11} +\lfloor \frac {n}2\rfloor E_{22}}
\frac {{\bf e}_{(\ell - 1 {\rm mod}\  2 )+1}}{\l^{\lfloor (\ell-1)/2\rfloor+1}} 
\label{mathbbG}
\eea
A similar computation yields the rest of formula \eqref{Gentries}.
Since we are interested in the determinant of $\mathbb G$ (up to multiplicative constants), we rearrange the basis in $W$; then the matrix can be written more transparently as 
\bea
\mathbb G_{k,\ell} =
\ds  \res{\lambda=\infty}  \frac{  \l^{\lfloor \frac{\ell -1}2\rfloor}\mathbf e_{\le\{2\atop 1\ri\}}^{\mathrm T}\G_0^{-1}(\le\{\l_k\atop \mu_{k-n_2}\ri\})G_{\infty}(\l){\mathbf e}_{(\ell - 1 \, {\rm mod}\, 2)+1 }}{(\l-\le\{\l_k\atop \mu_{k-n_2}\ri\})}  
\eea
where for brevity the notation $\{\}$ denotes two choices, according to the cases $k\leq n_2$ (top) or $k\geq n_2+1$ (bottom). 
  \paragraph{Variations of $\det \mathbb G$.}
It was shown in Appendix B of \cite{BertolaCafasso5}, Theorem B.1,  that (translating to the current setting) 
\be
\label{77}
\pa \ln \det \mathbb G =&\& \oint_{\pa \mathbb D} \Tr\le(\RR^{-1} \RR' \pa \J\J^{-1}\ri) \frac {\d\l}{2i\pi}  + \cr
&\& +\sum_{k=1}^{n_2} \oint_{\pa \mathbb D_k} \Tr\le(\G_0^{-1}\G_0' \pa (\l_k-\l)^{E_{22}} (\l_k-\l)^{-E_{22}}  \ri) \frac {\d \l }{2i\pi}  \cr
&\& +\sum_{k=1}^{n_1} \oint_{\pa \mathbb D_{n_2+k}} \Tr\le(\G_0^{-1}\G_0' \pa (\mu_k-\l)^{E_{11}} (\mu_k-\l)^{-E_{11}}  \ri) \frac {\d \l}{2i\pi} =\nonumber \\
=&\& \oint_{\pa \mathbb D} \Tr\le(\RR^{-1} \RR' \pa \J\J^{-1}\ri) \frac {\d\l}{2i\pi}  +\sum_{\zeta \in \vec \l, \vec \mu}
\res{\l=\zeta} \Tr \le(
\G_0^{-1}\G_0 ' \pa D D^{-1}\ri), 
\ee
where  $\pa$ means any  derivative of the $\l_k$'s, $ \mu_k$'s  or $x$ and in the last step we have used the fact that 
\be
\frac {\pa\sqrt{\l_k}}{\sqrt{\l_k}-\sqrt{\l} } = \frac {\pa\l_k}{\l_k-\l} + \mathcal O(1),\ \ \ \l\to \l_k,
\ee
and similarly for $\mu_k$.

\noindent {\bf Proof of Theorem \ref{taudressing}.}
We use the trivial algebra below with  $\G_n = R \G_0 D$ and $M_n = D^{-1} M D$:
\bea
\G_n^{-1}\G_n' &\& =  D^{-1}\G_0 ^{-1}  R^{-1} R' \G_0 D +  D^{-1}  \G_0^{-1} \G_0' D + D^{-1} D' ,
\nonumber\\
\pa (D^{-1} M  D) D^{-1}M ^{-1} D &\& =  D^{-1} \pa M  M ^{-1} D - \pa D D^{-1} + D^{-1}M D^{-1} \pa D M ^{-1} D,\nonumber\\
\label{ww}
\Gamma_{n,-} \pa M_{n}  M_{n} ^{-1} \Gamma_{n,-}^{-1} &\& =\pa  \Gamma_{n,+} \Gamma_{n,+}^{-1} - \pa \Gamma_{n,-} \Gamma_{n,-}^{-1}\ .
\eea
Plugging into the integrand of \eqref{Omega} and simplifying, using the cyclicity of the trace several times and \eqref{ww}, we find (below $\Delta$ denotes the jump operator $\Delta (f) = f_+ - f_-$):
\bea
&\Tr&\!\!\!\!\!\! \bigg[
\G_{n,-}^{-1}  {\G_{n,-}'}  \pa (D^{-1} M  D) D^{-1}M^{-1} D
\bigg]=\nonumber\\
&=& \Tr \bigg[
\Gamma_{0,-}^{-1} \Gamma_{0,-}' \pa M  M ^{-1} + 
R^{-1} R'  {\Delta \le(\pa (\G_0 D) D^{-1}\G_0^{-1}\ri)} + 
{ \Delta \le( \G_0^{-1} \G_0' \pa DD^{-1} \ri)}+\nonumber\\
& 
&{-  M ^{-1} M ' \pa D D^{-1}} + D^{-1} D' \le(  
\pa M  M ^{-1}  - \pa D D^{-1} + M D^{-1} \pa D M ^{-1} \ri)\bigg]. \label{modr}
\eea
Given the particular triangularity of the jump matrices $M $, all terms on the last line of \eqref{modr} are traceless and thus drop out. 

Now, if we have $\int_\Sigma \Delta F \frac{\d z}{2i\pi}$ and $F$ has some poles outside of $\Sigma$ then this reduces, by the Cauchy theorem, to the sum of the residues of $F$.  We are thus left with 
\bea
\Omega(\pa;[M_n])-\Omega(\pa;[M]) 
 =
 \int_\Sigma \Tr \le(
\G_{n}^{-1} \G_{n}' \pa M_{n} M_{n}^{-1}
\ri) \frac {\d \l}{2i\pi}
- \int_\Sigma \Tr \le(
\G_{0}^{-1} \G_{0}' \pa M M^{-1}
\ri) \frac {\d \l}{2i\pi}
=
\\
=
\oint_{\pa \mathbb D}  \Tr \le(
R^{-1} R' \pa (\G_0 D) D^{-1} \G_0^{-1} 
\ri) \frac {\d\l}{2i\pi} +\sum_{\z\in \vec \l, \vec \mu}
\res{\l=\zeta} \Tr \le(
\G_0^{-1}\G_0' \pa D D^{-1}\ri) 
=\\
=
\oint_{\pa \mathbb D}  \Tr \le(
\RR^{-1} \RR' \pa \wt \J \wt \J^{-1} 
\ri) \frac {\d\l}{2i\pi} +\sum_{\zeta \in \vec \l, \vec \mu}
\res{\l=\zeta} \Tr \le(
\G_0^{-1}(\l)\G_0 '(\l) \pa D(\l) D^{-1}(\l)\ri) 
\label{DeltaOmega} 
=\\
\mathop{=}^{\eqref{77}} \pa \ln \det \mathbb G +  \oint_{\pa \mathbb D}  \Tr \le(
\RR^{-1} \RR' (\pa \wt \J \wt \J^{-1}  -\pa  \J  \J^{-1} )
\ri) \frac {\d\l}{2i\pi}
\label{78}.
\eea
Note that we can substitute $R$ with $\RR$ in \eqref{DeltaOmega}, because these two matrices differ by a left multiplication with a $\l$--independent matrix, and the expression \eqref{DeltaOmega} is invariant under  this operation.
The matrix $\wt \J(\l)$ is read off the above formula and is given by
\be
\wt \J(\l)  = \le\{
\begin{array}{cc}
\G_0(\l) D(\l)   &  \l\in \pa \mathbb D_k\\
\G_0(\l) D(\l) \frac {\1 - i\s_1}{\sqrt{2}} \l^{\frac {\s_3}4} & \l \in  \pa \mathbb D_\infty.
\end{array}
\ri.
\ee
Therefore the matrix  $\wt \J$ differs from $\J$ \eqref{defJ} only on  the boundaries of the finite disks $\mathbb D_k,\ k=1,\dots, n$ by the factor  $T(\l)$ given by the diagonal matrix below
\be
\wt \J(\l)  = \J(\l) T(\l)\ ,\qquad
T(\l) = D(\l) {\prod_{k=1}^{n_2}  (\l_k - \l)^{- E_{22} \chi_{_{\mathbb D_k}} }} {\prod_{k=1}^{n_1}  (\mu_k - \l)^{- E_{11} \chi_{_{\mathbb D_{n_2+k}}} }},
\ee
where $\chi_{_X}$ denotes the indicator function of the set $X$. 
Note that $T(\l)$  belongs to $\H_+(\mathbb D_k), \ \forall k$. 
We now  follow the exact same steps as in the proof of Theorem B.2 of \cite{BertolaCafasso5} and have 
\bea
 \oint_{\pa \mathbb D}&\& \Tr\le(\RR^{-1} \RR' (  \pa \wt \J \wt \J^{-1} -\pa \J  \J^{-1} ) \ri) \frac {\d\l}{2i\pi} =\nonumber \\
 =&\&
  -\sum_{k=1}^{n_2} \res {\l=\l_k} \Tr \le(\frac {E_{22} }{\l-\l_k} \pa T T^{-1}  \ri) 
  -\sum_{k=1}^{n_1} \res {\l=\mu_k} \Tr \le(\frac {E_{11} }{\l-\mu_k} \pa T T^{-1}  \ri)  =\cr
 =&\&
 \sum_{k=1}^{n_2}\le(
 \frac  {\pa \l_k}  {4\l_k}-
  \sum_{j\neq k}  \frac {\pa \sqrt{\l_j}}{\sqrt{\l_j}-\sqrt{\l_k}}\ri)
  +
 \sum_{k=1}^{n_1}\le(
 \frac  {\pa \mu_k}  {4\mu_k}-
  \sum_{j\neq k}  \frac {\pa \sqrt{\mu_j}}{\sqrt{\mu_j}-\sqrt{\mu_k}}\ri)
  - \sum_{k=1}^{n_2} \sum_{j=1}^{n_1} \frac {\pa (\sqrt{\mu_j} + \sqrt{\l_k})}{\sqrt{\mu_j} + \sqrt{\l_k} }
 \nonumber\\
=&\&   \pa \ln
\frac{\ds \prod_{j=1}^{n_2} {\l_j}^\frac 1 4   \prod_{j=1}^{n_1} {\mu_j}^\frac 1 4}
{ \ds  \prod_{j<k\leq n_2} (\sqrt{\l_j} - \sqrt{\l_k})\prod_{j<k\leq n_1} (\sqrt{\mu_j} - \sqrt{\mu_k})   \prod_{j=1}^{n_1}\prod_{k=1}^{n_2} (\sqrt{\mu_j} + \sqrt{\l_k}) 
 } =: \pa \ln \Delta(\vec \l , \vec \mu). \label{133}
\eea
Combining \eqref{133} with \eqref{78} we obtain
\bea
&\& \Omega(\pa;[M_n])-\Omega(\pa;[M]) 
 \mathop{=}^{\eqref{78}}
 \pa \ln \det \mathbb G +  \oint_{\pa \mathbb D}  \Tr \le(
\RR^{-1} \RR' (\pa \wt \J \wt \J^{-1}  -\pa  \J  \J^{-1} )
\ri) \frac {\d\l}{2i\pi}\mathop{=}^{\eqref{77}}\cr
&\& =
\pa \ln \le( {\det \mathbb G}\; \Delta(\vec \l,\vec \mu)\ri)
\eea
The proof is complete.
\QED
\section{Explicit computation of $Z_n$}
\label{proofZnKY}
For the benefit of the reader we derive \eqref{ZnKY}  (in a way that is slightly different from \cite{Kontsevich:1992p30}) as follows.
Let  $\d U$ be the  Haar measure on $U(n)$, $S = {\rm diag} (s_1,\dots, s_n)$ and the $s_j$'s  the eigenvalues of $M$, and $\Delta(S) = \prod_{j<k} (s_j-s_k)$.
Considering the numerator of \eqref{ZnK}, and  setting $\Lambda =  {Y}^2$ we have
\bea
\int_{H_n} &\& \d M {\rm e}^{ \Tr \le(i\frac {M^3}3 -  {Y} M^2  + i x M \ri)}
\mathop{=}^\star
{\rm e}^{\frac 2 3 \Tr  {Y}^3 + x \Tr Y} \int_{H_n} \d M {\rm e}^{i \Tr \le(\frac {M ^3}3 +  ({Y}^2 + x )M\ri)} \mathop{{=}}^{\hbox{\tiny Weyl integration formula}}
\cr
&\&{=C_n}
{\rm e}^{\frac 2 3 \Tr  {Y}^3 + x \Tr Y}\int_{\R^n}\Delta^2(S) \prod_{j=1}^{n} {\rm e}^{\frac {is_j^3}{3} + i s_j x } \d s_j \int_{U(n)} \d U{\rm e}^{i \Tr( {\Lambda} U S U^\dagger)} \mathop{=}^{\tiny \hbox{(Harish-Chandra)}}
\cr
&\& {=\wt C_n} {\rm e}^{\frac 2 3 \Tr  {Y}^3 + x \Tr Y }\int_{\R^n}  \frac {\Delta(S)\det\le[{\rm e}^{i s_j \l_k}\ri]_{j,k\leq n}}{\Delta( {\Lambda})}\prod_{j=1}^{n} {\rm e}^{\frac {is_j^3}{3}+ is_j x } \d s_j\mathop{=}^{\tiny \hbox{Andreief}}
\cr
&\&=
\frac{ \wt C_n n! {\rm e}^{\frac 2 3 \Tr  {Y}^3 +  x \Tr Y} }{\Delta( {\Lambda})}\det\le[\int_{\R}s^{j-1} {\rm e}^{\frac {i s^3}3 + i s (\l_k+ x )} \d s \ri]_{j,k\leq n}
=\\
&\&= 
{\wt C_n n! (2\pi)^n{\rm e}^{\frac 2 3 \Tr  {Y}^3+ x \Tr Y} }\frac{\det\le[\Ai^{(j-1)} (\l_k+ x ) \ri]_{j,k\leq n}}{\Delta( {\Lambda})}\nonumber
\label{Qn}
\eea
where $C_n, \wt C_n$   are proportionality constants (depending only on $n$) of no present interest (it turns out that $\wt C_n = \frac {\pi^{\frac n2(n-1)}} {n!}$).
In the step marked with $\star$ we have performed  a shift $M \mapsto M-iY$ and an analytic continuation; the integral is now only conditionally convergent and it can be understood as absolutely convergent integration on $H_n + i\epsilon \1,\ \epsilon>0$.

Recall now that 
\be
\int_{H_n} \d M {\rm e}^{- \Tr \le( {Y} M^2\ri)}  =
\frac{ {\pi ^ {\frac n 2+ \frac n 2 (n-1) } } }{\prod_{j=1}^n\sqrt{ y_j}\prod_{j<k}^{n} (y_j+y_k) } =  \frac{\pi^{{\frac {n^2}2}} }{
 \prod_{j=1}^{n} \l_j^{\frac 1 4}\prod_{j < k=1}^{n} {\sqrt{\l_j}+\sqrt{\l_k}} }.
\ee

Thus, in total 
\be
Z_n({Y}) = 2^n \pi^\frac n2 {\rm e}^{\frac 2 3 \Tr  {\Lambda}^\frac 32  + x  \Tr \sqrt{\Lambda}}\frac{\det\le[\Ai^{(j-1)} (\l_k+ x ) \ri]_{j,k\leq n} \prod_{j=1}^n(\l_j)^{\frac 1 4}  \prod_{j<k}(\sqrt{\l_j}  + \sqrt{\l_k})}{\Delta( {\Lambda}) }.\
\ee 
The overall proportionality  constant  is determined by observing that $Z_n$ as defined in \eqref{ZnK} tends to $1$  as the eigenvalues of $Y$ tend all to $+\infty$. The Airy function has the asymptotic behavior
\be
\Ai(\l) =\frac{{\rm e}^{-\frac 2 3 \l^{\frac 32}}}{2\sqrt{\pi} \l^{\frac 1 4}}( 1 + \mathcal O(\l^{-\frac 3 2}))\ ,\ \ |\arg(\l)|<\pi,
\ee
and hence 
\be
\det\le[\Ai^{(j-1)} (y_k^2+ x ) \ri]_{j,k\leq n}  \simeq \frac {{\rm e}^{-\sum_{j=1}^{n} \le(\frac 2 3 y_j^3  +  x y_j \ri)}}{2^n \pi^{\frac n2} \prod_{j=1}^{n} y_j^\frac 1 2} \prod_{j<k} (y_j - y_k)\qquad \hbox{ as } x\to\infty
\ee
from which the proportionality constant is deduced. The expression \eqref{ZnKY} follows from \eqref{Qn}  by substituting $\Lambda = Y^2$ and simplifying.

\br
\label{REMB1}
The above computation is carried out under the assumption  that $\Re y_j>0$; to see what happens when $\Re y<0$, 
 consider the simplest case $n=1$ and $Y = y<0$; then (set $x=0$ for simplicity)
\be
\frac{ \int_\ell \d M {\rm e}^{i\frac {M^3}3 - Y M^2}}{\int_\ell {\rm d}M{\rm e}^{-YM^2}} = {\rm e}^{ \frac 23y ^3 } \int_\ell {\rm e}^{i\frac{ s^3}3 +i s y^2} \d s
\label{n1}
\ee
In order to be able to interpret \eqref{n1} as an average of ${\rm e}^{i\frac {M^3}3}$ with respect to a Gaussian measure, we must choose the path of integration $\ell$ in both numerator/denominator so as to have a convergent integral and also so that the term ${\rm e}^{i\frac {M^3}3}$ is oscillatory.  The choice $\arg(s) = {\rm e}^{\pm i \frac \pi 3}$ is possible. But this means that the integral gives now $\Ai(\omega^{\pm1}  y^2)$ rather than $\Ai(y^2)$. This is the underlying reason for the definition \eqref{ZnKcont}
\er

\section{Proof of Prop. \ref{propdetG}}
\label{belinproof}
 Denote the column vectors  $G_\infty(\l) {\mathbf e}_{1,2}$  (with $G_\infty$ as in \eqref{Ginfty}) by $H_{1,2}(\l)$ (respectively) and the row vectors 
 \be
 \label{lammu}
\A_k:=  \le\{
{\mathbf e}_{2}^{\mathrm T} \G_0^{-1}(\l_k) \ \  \ \ \ \ \ \  k\leq n_2 \ \ \ \ \ \  \ \ \ \ \ \ 
\atop 
{\mathbf e}_{1}^{\mathrm T} \G_0^{-1}(\mu_{k-n_2}) \ \ \ \ n_2+1\leq k \leq n_1+n_2.
\ri.
\ee
The explicit expression depends on which sector $\l_k$'s, $\mu_j$'s belong to, and can be read off from \eqref{defA}.
Consider the wedge of the first two columns of $\mathbb G$;
\be
\le[ \A_k \res{\lambda=\infty}  \frac{ H_1}{(\l-\l_k)} \ri]_{k=1}^n\wedge   \le[\A_k \res{\lambda=\infty}  \frac{ H_2}{(\l-\l_k)} \ri]_{k=1}^{n}.
\ee
Here and below, $[...]_{k=1}^n$ denotes  column vectors indexed by $k$.
Depending on the parity of $n$ and using \eqref{Jinfty0}, \eqref{Jinfty},  we have (the symbol $\star$ denotes a constant that eventually drops out of the computation, hence irrelevant)
\bea
\res{\l=\infty} \frac {H_1}{\l-\l_k}  =\res{\l=\infty} \frac {H_1}{\l}  =\le\{\begin{array}{cc}
 {\bf e}_1 + \star {\bf e}_2 & n\equiv 0 \ {\rm mod} \, 2,\\
 -i{\bf e}_2& n\equiv 1 \ {\rm mod} \, 2,\\
 \end{array}
 \ri.\nonumber 
\\
 \res{\l=\infty} \frac {H_2}{\l-\l_k} =\res{\l=\infty} \frac {H_2}{\l}  =\le\{\begin{array}{cc}
 {\bf e}_2& n\equiv 0 \ {\rm mod} \, 2,\\
i{\bf e}_1 + \star {\bf e}_2& n\equiv 1 \ {\rm mod} \, 2.\\
 \end{array}
 \ri.
 \label{H12}
\eea
Therefore, for any parity of $n$ (up to an inessential sign), we have 
\bea
\bigg[ \A_k \res{\lambda=\infty}&\& \frac{ H_1}{(\l-\l_k)} \bigg]_{k=1}^n\wedge   \le[\A_k \res{\lambda=\infty}  \frac{ H_2}{(\l-\l_k)} \ri]_{k=1}^{n}
 =\pm 
\le[ (\A_k)_{1}  \ri]_{k=1}^n\wedge   \le[(\A_k)_2  \ri]_{k=1}^{n}.
\label{C4}
\eea
Consider now the third and fourth columns: for the third one we have 
\bea
\le[ \A_k \res{\lambda=\infty}  \frac{  \l H_1}{(\l-\l_k)} \ri]_{k=1}^n &\& 
=\le[ \l_k \A_k \res{\lambda=\infty}  \frac{  H_1}{(\l-\l_k)} \ri]_{k=1}^n  + \le[ \A_k \res{\lambda=\infty}  \frac{  (\l-\l_k) H_1}{(\l-\l_k)} \ri]_{k=1}^n \nonumber \\
&\&=\le[ \l_k \A_k \res{\lambda=\infty}  \frac{  H_1}{\l} \ri]_{k=1}^n  + \le[ \A_k \res{\lambda=\infty}   H_1 \ri]_{k=1}^n 
\label{3rd}
\eea
and similarly for the fourth
\bea
\le[ \A_k \res{\lambda=\infty}  \frac{  \l H_2}{(\l-\l_k)} \ri]_{k=1}^n =\le[ \l_k \A_k \res{\lambda=\infty}  \frac{  H_2}{\l} \ri]_{k=1}^n  + \le[ \A_k \res{\lambda=\infty}   H_2\ri]_{k=1}^n.
\label{4th} 
\eea
The last terms in \eqref{3rd}, \eqref{4th}, respectively,  are in the span on the first two columns appearing in \eqref{C4} and hence can be dropped. The first two term are in the span of 
\be
\le[ \l_k(\A_k)_{1}  \ri]_{k=1}^n\wedge   \le[\l_k(\A_k)_2  \ri]_{k=1}^{n}.
\ee
Proceeding this way by induction we arrive at 
%
\be
\det \mathbb G = \bigwedge _{r=1}^{n} \le[ \l_k^{\le\lfloor \frac {r-1}2\ri\rfloor} \A_k \res{\lambda=\infty}  \frac{  H_{2 - (r \,{\rm mod}\, 2)}}{\l} \ri]_{k=1}^n.  
\ee
To have more compact formu\ae\ we denote $f_k:= \AA_{s_k}(\l_k + x)$ and $g_k:= \AA_{s_k} (\mu_k+x)$, where $s_k$ is the sector to which $\l_k$ (or $\mu_k$) belongs  as indicated in the statement of the proposition and which follows from \eqref{defA} and \eqref{lammu}. Then, depending on the parity of $n$,  we have explicitly,  for even $n$:
\bea
\det &\&\mathbb G = (-2i\pi)^{\frac n2}{\rm e}^{\le(
 \sum_{j=1}^{n_2} \le(\frac 2 3\l_j^\frac 32 +  x\l_j^\frac 1 2\ri) -\sum_{\ell=1}^{n_1} \le(\frac 2 3\mu_\ell^\frac 32 +  x\mu_\ell^\frac 1 2\ri)\ri)}
\det 
\le[
\le[
f_k
  \bigg| 
 f'_k
\bigg| \cdots \bigg| 
 \l_k^{\frac n2} 
f_k
 \bigg| 
\l_k^{\frac n2} 
f'_k
\ri]_{k=1}^{n_2} 
\atop
\le[
g_k
 \bigg| 
g_k'
\bigg| \cdots \bigg| 
 \mu_k^{\frac n2} 
g_k
 \bigg| 
\mu_k^{\frac n2} 
g_k'
\ri]_{k=1}^{n_1} 
\ri]
\nonumber
\eea
and, for odd $n$,
\bea
&\& \det \mathbb G = (-2i\pi)^{\frac n2} {\rm e}^{\le(
 \sum_{j=1}^{n_2} \le(\frac 2 3\l_j^\frac 32 +  x\l_j^\frac 1 2\ri) -\sum_{\ell=1}^{n_1} \le(\frac 2 3\mu_\ell^\frac 32 +  x\mu_\ell^\frac 1 2\ri)\ri)}
\det 
\le[
\le[
f_k
\bigg| 
f_k'
\bigg| 
\cdots \bigg| 
\l_k^{\frac {n-1}2}  
f_k
\bigg|
\l_k^{\frac {n-1}2}  
f_k'
 \bigg|
\l_k^{\frac {n+1}2}
f_k
\ri]_{k=1}^{n_2}
\atop
\le[
g_k  
\bigg| 
g_k'
\bigg| 
\cdots \bigg| 
\mu_k^{\frac {n-1}2}  g_k
\bigg|
 \mu_k^{\frac {n-1}2}  g_k'
 \bigg|
\mu_k^{\frac {n+1}2} g_k
\ri]_{k=1}^{n_2}
\ri].
\nonumber
\eea
Using the differential equation for the Airy functions repeatedly (all $f_k, g_k$'s solve the same Airy  ODE), and further elementary (triangular) column operations, we see easily that in all cases (up to an inessential sign) we obtain the formula \eqref{GnK}. \QED

\bibliographystyle{plain}
\def\cprime{$'$}

\end{document}